\documentclass[aastex]{emulateapj}

\begin{document}

\title{Large-Scale Image Processing with the ROTSE Pipeline for Follow-Up of 
Gravitational Wave Events}
 
\author{
L.~K.~Nuttall$^{1}$,
D.~J.~White$^{2}$,
P.~J.~Sutton$^{1}$,
E.~J.~Daw$^{2}$,
V.~S.~Dhillon$^{2}$,
W.~Zheng$^{3}$,
C.~Akerlof$^{4}$
}
\affiliation{$^{1}$School of Physics and Astronomy, Cardiff University, Cardiff, CF24 3AA, United Kingdom}
\affiliation{$^{2}$Department of Physics and Astronomy, University of Sheffield, Sheffield, S3 7RH, United Kingdom}
\affiliation{$^{3}$Department of Astronomy, University of California, Berkeley, CA 94720-3411, USA}
\affiliation{$^{4}$Randall Laboratory of Physics, University of Michigan, 450 Church Street, Ann Arbor, MI, 48109-1040, USA}


\begin{abstract}
Electromagnetic (EM) observations of gravitational-wave (GW) sources would 
bring unique insights which are not available from either channel alone. 
However EM follow-up of GW events presents new challenges. GW events will have 
large sky error regions, on the order of 10-100 deg$^{2}$. Therefore there is 
potential for contamination by EM transients unrelated to the GW event. 
Furthermore, the characteristics of possible EM counterparts are uncertain, 
making it desirable assess the statistical significance of a candidate EM 
counterpart. Current image processing pipelines are not usually optimised for 
large-scale processing. We have automated the ROTSE image analysis, and 
supplemented it with a post-processing unit for candidate validation and 
classification. We also propose a simple \textit{ad hoc} statistic for ranking 
candidates as more likely to be associated with the GW trigger. We demonstrate 
the performance of the automated pipeline and ranking statistic using archival 
ROTSE data. EM candidates from a randomly selected set of images are compared 
to a background estimated from the analysis of 102 additional sets of archival 
images. The pipeline's detection efficiency is computed empirically by 
re-analysis of the images after adding simulated optical transients that follow typical lightcurves for gamma-ray burst afterglows and kilonovae. The automated 
pipeline rejects most background events, and has $\simeq$50\% detection 
efficiency for transients up to the real limiting magnitude of the images. 
However $\sim$10\% of the image sets show a residual background tail that 
impedes assigning a high significance to any putative candidate. This motivates 
the use of information beyond simple lightcurves for background rejection.
\end{abstract}



\section{Introduction}

Multi-wavelength and multi-channel observations of astrophysical systems can 
yield insights in to the system that are not available from a single waveband. 
For example, the detection of gamma-ray burst (GRB) systems in the x-ray, 
optical and radio bands have led to the identification of host galaxies and 
their redshifts, in addition to tests of theoretical models 
\citep{Bloom:2005qx,Soderberg:2006bn,nakar:2007}. 
Similar benefits may be expected from 
multi-channel follow up of systems that emit gravitational waves (GWs). Some of 
the anticipated advantages include identifying host galaxies, improving 
parameter estimation of GW events, and determining the progenitors of 
phenomena such as short hard gamma-ray bursts (SGRBs); see for example 
\citet{Bloom:2009vx}. 
The first attempts to detect electromagnetic (EM) counterparts to candidate GW 
events were made during the 2009-2010 science run of the LIGO and Virgo 
detectors 
\citep{virgo:2012,Abbott:2007kv}.
Details on how this search was performed are documented in 
\citet{Abbott:2011ys}. 
A number of optical telescopes were triggered by the GW detectors, one 
such system being ROTSE-III. Given the GW detector sensitivities at the time 
of the search, it is unlikely that any of those triggers represent true 
astrophysical events. However these joint observations are a useful 
exercise in preparing for the era of Advanced gravitational-wave detectors 
\citep[][c. 2015+]{adv,aligo,Aasi:2013wya}, when EM follow-ups will be performed on GW 
triggers of astrophysical origin.

The ROTSE collaboration has a well established image processing pipeline. This 
pipeline makes use of astronomical image subtraction by cross-convolution, 
removing the need for high quality reference images, with similar computational 
efficiency to other image processing procedures \citep{Yuan:2008ur}. Transient 
identification is based on human scanning of potential candidates identified by 
the pipeline, and separate generation of lightcurves of the most interesting 
candidates. The pipeline has proven to be successful in finding supernovae as 
well as GRB afterglows etc. \citep{Rykoff:2005es,Quimby:2007gv,Rykoff:2009ma}. 
However, the detection of optical transients associated to GW triggers presents 
new challenges, in particular the need to process large numbers of images to 
cover a typical GW error region, and the ability to assign a quantitative false 
alarm probability on any detected optical transient. It is therefore essential 
that we have an automated image processing pipeline, where large numbers of 
images can be processed. 

In this paper we present modifications made to the ROTSE 
pipeline to allow the processing of large numbers of images with automated 
detection and tentative classification of transients. We evaluate the 
performance using archival ROTSE images, and use custom-built software to add 
simulated transients to images. The paper is organised as follows. In Section 
\ref{sec:challenges} we discuss the challenges associated with detecting an EM 
counterpart to a GW event. In Section \ref{sec:rotse} we give a brief summary 
of the ROTSE-III telescope system as well as the images ROTSE took. 
In Section \ref{sec:pipeline} we summarise how the ROTSE image 
processing pipeline identifies candidates. In Section 
\ref{sec:mods} we describe the modifications made to automate the pipeline, 
including details of how the most significant candidates are identified as well 
as the simulation procedure. In Sections \ref{sec:background} and 
\ref{sec:injperf} we discuss the results of processing archival images to 
evaluate the optical transient background, and processing simulated transients 
to quantify the performance of the pipeline. 
We conclude with some brief comments in Section \ref{sec:concl}.

\section{Detecting an EM counterpart of a GW Event}
\label{sec:challenges}

Many systems which produce detectable GWs should also be observable in EM 
wavebands \citep{Abbott:2011ys}.
The most promising GW sources which are also expected to have EM 
counterparts are mergers of binary neutron stars (NS-NS) or binaries consisting 
of a neutron star and stellar mass black hole (NS-BH). These systems are also 
the favoured progenitor model for SGRBs \citep{nakar:2007}. 
\citet{Abadie:2010cf} and \citet{Aasi:2013wya} 
summarise predictions of the rate of detection of such systems by the Advanced 
LIGO detectors. \citet{MeBe:12} review various possible EM counterparts. In 
addition to SGRBs, these include orphan optical/radio afterglows, and 
supernova-like optical or near-IR transients (`kilonovae') generated by the decay of 
heavy nuclei produced in the merger ejecta \citep{Li:1998bw,Metzger:2010sy}. 
Another system which may produce detectable GWs are long gamma-ray bursts 
(LGRBs); see \citet{s6grb} for a summary of possible GW emission scenarios.
There is a wealth of 
observational data detailing the afterglow of both SGRBs and LGRBs.
Observations detailed in \citet{Kann:2007cc,Kann:2008zg} indicate that one day 
after detection, the afterglow magnitude will be in the range 18-24 for a LGRB 
and 24-30 for a SGRB (for a source at $z=1$) and follow a power-law decay 
constant of approximately -2.6.
The optical kilonova transient is expected to produce an optical emission peak
at magnitude 18 at one day for a source at 50 Mpc and fade over the course of a
few days \citep{Metzger:2010sy}.

GW events which produce high-energy EM counterparts such as GRBs may 
be promptly identified and localised by satellites such as Swift \citep{swift} 
and Fermi \citep{fermi}. However, for GW events where high-energy emission is 
absent, or beamed away from Earth, or where the source is outside the field of 
view of these satellites, the detection of an EM counterpart to a GW event will 
be challenging. First, sky localisation using a GW data alone will produce a 
large error box, typically 10 -- 100 deg$^{2}$ 
\citep{Fairhurst:2009tc,Fairhurst:2010is}. The field of view (FOV) of one 
of the ROTSE-III telescopes is $\sim3$ deg$^{2}$, making it impractical to 
image the entire error region. Instead, we make use of the fact that current GW 
detectors had a maximum distance sensitivity of between 30-70 Mpc (for NS-NS 
and NS-BH binary mergers) \citep{Abadie:2010cf} 
and focus observations upon galaxies in the error region within the reach of 
GW detectors using the galaxy catalogue described in 
\citet{White:2011qf}. 
Despite there being hundreds of galaxies in a 
typical GW error box, the galaxies can be ranked according to their distance 
and luminosity as the most likely host from which the signal originated. 
Considering a typical pointing with a ROTSE-III telescope, the probability of 
successfully imaging the correct host galaxy is estimated at between 30\%-60\%, 
not including galaxy catalogue incompleteness \citep{Nuttall:2010nk}. For the 
Advanced GW detectors, which will have an order of magnitude larger distance 
reach \citep{adv,aligo}, preliminary estimates indicate that at least $\sim$10 
pointings will be required to have reasonable probability of imaging the host 
galaxy \citep{nuttall2}. \citet{Nissanke:2012dj} also present 
strategies for identifying EM counterparts to GW mergers in the Advanced 
detector era.

Another complication of detecting EM counterparts to GW events is that 
the magnitude and decay timescale of possible EM counterparts are uncertain 
\citep{Abbott:2011ys}. 
This uncertainty necessitates observations at both early and late times, 
ideally from seconds to weeks after the trigger. Combined with the large 
error regions associated with GW triggers, this implies the need to process 
many images. Given the uncertain nature of the counterpart lightcurve, the 
image analysis should be capable of detecting any transient 
that is inconsistent with typical background events (which may be real 
astrophysical transients unrelated to the GW trigger or image artefacts).

Finally, there has not been a confirmed detection of a GW to date, making it 
desirable to be able to assign a high statistical confidence in any putative EM 
counterpart. Analysing both `background' images (images from pointings 
not associated with a GW trigger) and `injection' images (images containing 
simulated transients with known lightcurves) 
will be vital to quantify the rate at which false transients 
are detected as well as the performance of the pipeline. In particular, we need 
to test any background rejection steps on injected transients to verify they 
are `safe'. All of these factors point to the need to automate the EM image 
analysis \citep[see for example][]{Bloom:2011as} to allow 
large-scale processing and quantitative characterisation of the pipeline. 

\section{The ROTSE-III Telescope System}
\label{sec:rotse}

The Robotic Optical Transient Search Experiment (ROTSE) is dedicated to rapid 
follow up observations of GRBs and other fast optical transients on the time 
scale of seconds to days. ROTSE has undergone two phases of development thus 
far, ROTSE-I 
and III. ROTSE-I consisted of a 2 x 2 array of telephoto camera lenses co-
mounted on a rapid-slewing platform, located in northern New Mexico. The array 
was fully automated and started taking data in 1998. Observations made by 
ROTSE-I of GRB 990123 revealed the first detection of an optical burst 
occurring during the gamma-ray emission, demonstrating the value of autonomous 
robotic telescope systems \citep{Kehoe:1999zc}.

The ROTSE-III telescope system came online in 2003 and consists of four 0.45m 
robotic reflecting telescopes located in New South Wales, Australia (ROTSE-
IIIa), Texas, USA (ROTSE-IIIb), Namibia (ROTSE-IIIc) and Turkey (ROTSE-IIId). 
The instruments are fully automated and make use of fast optics to give a 1.85 
$\times$ 1.85 degree FOV. 
Under ideal conditions, ROTSE-III is capable of attaining 17th magnitude 
at the center of the FOV 
with a 5 second exposure, and 18.5 magnitude with a 60 second exposure. If 
multiple images are stacked on top of one another or `coadded' ROTSE-III can 
reach $\sim$19th magnitude \citep{Smith:2002bd}. 
The typical limiting magnitude away from the center of the FOV is 14; 
this is the appropriate measure of sensitivity when searching for transients
over the full 1.85 $\times$ 1.85 degree FOV.

Between September 2 and October 20 2010, ROTSE-III took over 700 images in 
response to 5 candidate GW triggers as part of the latest science run of the 
LIGO and Virgo detectors \citep{Abbott:2011ys}. All four ROTSE telescopes 
were used to gather the images, which span from the first night following the 
event to one month later and vary in exposure length (either 20 or 60 
seconds). When a LIGO-Virgo trigger was sent to the ROTSE telescopes, typically 
30 images were taken on the first night and 8 images taken on subsequent 
follow-up nights, per telescope, for the first ten nights following the 
trigger, with additional observations around nights 15 and 30. While we do not 
use these images in this paper, we use archival images selected with this 
cadence so as to characterise the automated ROTSE pipeline in conditions 
matching those of GW follow-up observations.

\section{The ROTSE Image Processing Pipeline}
\label{sec:pipeline}

\subsection{Basic features}
\label{subsec:works}

The ROTSE image processing pipeline \citep{Yuan:2008ur} was developed by the 
ROTSE collaboration 
to search for transient objects in images taken with the ROTSE-III telescopes. 
The pipeline makes use of cross-convolution to perform image subtraction. 
Image subtraction is an essential tool needed to remove contributions 
from static sources and amplify any subtle changes. For example, without 
image subtraction it would be almost impossible to find a source buried within 
a host galaxy. In this section we give a brief summary of the pipeline; 
more details can be found in \citet{Yuan:2008ur}.

The pipeline starts by processing images through \textsc{SExtractor} 
\citep{Bertin:1996aa}, giving a list of objects with precise 
stellar coordinates. These coordinates are used to compute corrections for 
image warping, so that the stellar objects within the image overlay as 
closely as possible with those in the reference image. It is essential 
to use an image or stacked set of images (see Section \ref{subsec:tocoadd}) 
of the same region from an uninteresting time as the reference image so that a 
new transient may be identified. At this point in the analysis 
pixels within both images which exceed the saturation level are 
excluded. To estimate the background as precisely as possible the background 
difference is found between the two images, instead of the individual 
background for each image separately. The sky difference map is generated 
by performing a pixel-by-pixel subtraction between the warped and the 
reference image and it is this which is subtracted from the original image. 
The main benefit of this sky difference map is that the final subtracted 
image will be background-free. This procedure is repeated for all images which 
are to be processed before the cross-convolution algorithm is invoked.

\subsection{Coadding}
\label{subsec:tocoadd}

On a typical night, two sets of four images of 60 second exposure\footnote{
A 20 second exposure is used if the target is in the vicinity of a bright 
galaxy or if the moon is in a bright phase.} with a 30 minute cadence are 
taken. These images are of the same 
part of the sky, so that images may be stacked on top of one another or 
`coadded'. Coadding increases, by about one magnitude, the limiting 
magnitude to which we are sensitive, allowing fainter objects to be seen 
without saturating the brightest objects within the image. Each 
four-image set is coadded, as well as the eight 
images taken for the night, resulting in three co-additions. These three 
images are then subtracted from the same reference image, 
and the three difference images processed through \textsc{SExtractor} to 
reveal the residual objects.

The ROTSE pipeline can also perform a `non-coadded' analysis, in which just 
the images taken from the first night are processed without coadding to see if 
there are any fast transients on the hour time scale. Since the non-coadded 
analysis does not stack images, the images have a shallower limiting 
magnitude than those images which have been coadded. In this paper we present
examples using the coadded method only, i.e. characterising the ability to 
detect transients with a characteristic timescale of a few days.

\subsection{Candidate Selection}
\label{subsec:cand}

In the coadded analysis, we have two images made from two sets of four images 
(called hereafter the `4-fold images') 
and one image made from the coadditions of all the images taken over the night 
(the `8-fold image') as described in Section \ref{subsec:tocoadd}. 
Any residual objects identified in these images by the pipeline 
are required to fulfil certain criteria to be considered candidate 
transients, as detailed in \citet{Yuan:2010ff}. 
First, the object must have a signal-to-noise ratio (SNR) above 
2.5 in the 4-fold images and above 5 in the 8-fold image. Next, 
the position of the object between the 4-fold and 8-fold images 
must match to within 1.5 pixels for candidates with SNR $<$ 15 
and to within 1 pixel for objects with SNR $>$ 15. 
The full width half maximum (FWHM) of the object must 
be no bigger than twice the median FWHM of the stars in the convolved 
reference image, as well as be within the range of one pixel.
The change in flux is also checked in a circular region of diameter 8 pixels 
around the object. Different cuts are applied depending 
on whether the potential candidate corresponds to a stellar object or lies in 
a known galaxy. For example, if an object matches a star 
or an unknown object a flux change of 60\% is required, whereas if the object 
is within 20\% of the semi-major axis length from the 
galaxy centre, but not consistent with a core, only a 3\% flux change is 
required \citep{Yuan:2010ff}.

After the potential candidates have gone through these checks,
further criteria are applied should more than twenty 
candidates remain. So many candidates remaining may indicate that 
the subtraction did not work correctly, or that the image 
quality is poor. First source crowding is checked, wherein potential 
candidates are rejected if they have more than 15 other potential 
candidates with 250 pixels. If there are still more than 20 potential 
candidates remaining, objects near the edge of the image are 
discarded, since the edges are liable to fringing and aberrations 
\citep{Yuan:2010ff}. Again, if more than 20 potential candidates remain, 
the area is reduced and the process repeated until the area of the image is 
800 pixels in width or there are less than 20 potential candidates 
remaining. In these situations it is not very likely that something of 
astrophysical significance will be found due to the quality 
of the images.

Objects which have passed all the criteria outlined above form the candidate 
list. In fact, several candidate lists are generated: one for each night 
in the coadded case, and one for each consecutive pair of images in the 
non-coadded case. These lists need to be combined to produce a single list of 
unique candidates. 
The vast majority ($\sim95\%$) of these potential candidates will be 
image subtraction artefacts, with a minority ($\sim2\%$) due to known variable 
objects such as variable stars or asteroids. We identify and remove these known 
transients by comparing to the SIMBAD catalogue\footnote{http://simbad.u-strasbg.fr/simbad/} and the 
Minor Planet Checker\footnote{http://scully.cfa.harvard.edu/cgi-bin/checkmp.cgi}. 

\subsection{Webpages}
\label{subsec:webpage}

\begin{figure}
\centering
\includegraphics[width=0.45\textwidth]{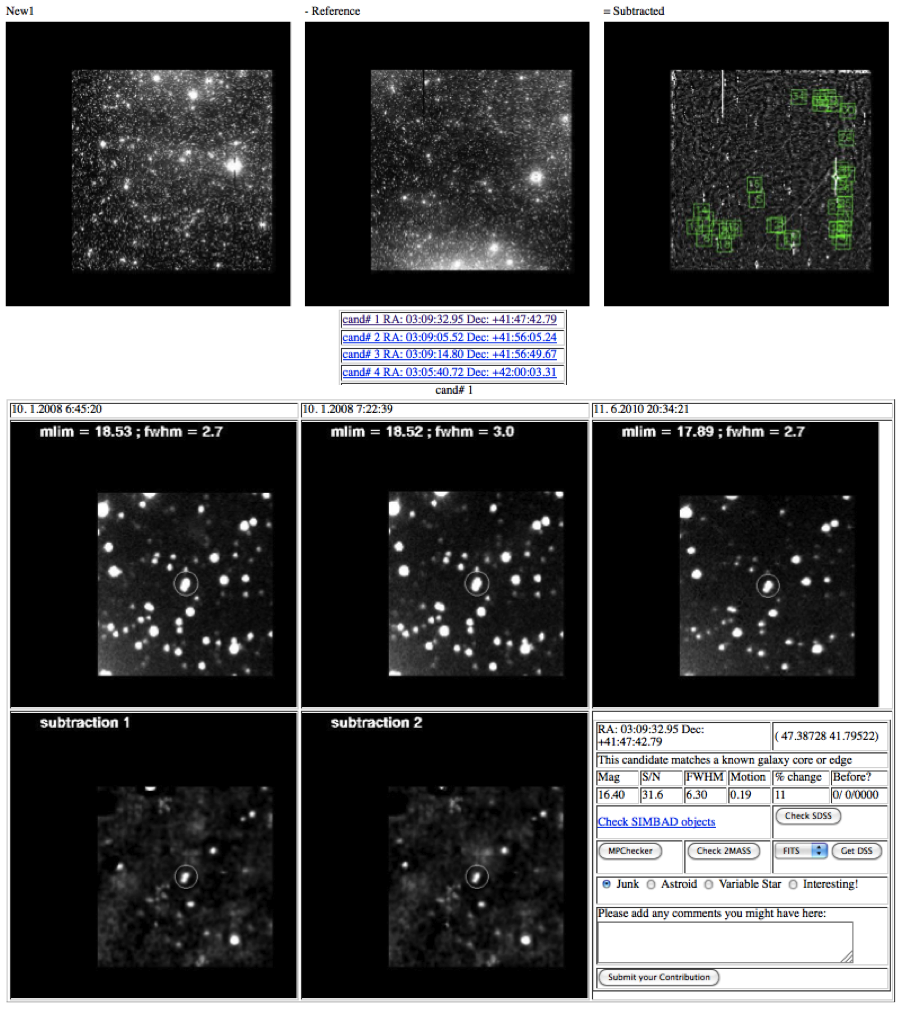}
\caption{\label{fig:webpages} A sample ROTSE pipeline webpage, showing links to 
all the candidates found as well as a table displaying subimages and 
information for the first candidate. The full webpage displays one table 
for each candidate.}
\end{figure}

For each candidate list the pipeline also generates 
a webpage such as the one shown in Figure \ref{fig:webpages}. At the top of 
the webpage three images are shown. On the left is the coadded image for one 
night, in the middle is the reference image, and on the right is the subtracted 
image. The example subtracted image shows four candidates. Below this are a 
list of links, one for each candidate. Selecting a link (in this case the 
first) displays a table of sub-images for that candidate. The top left panel of 
this table shows the first coadded image (from images 1-4 taken on that 
night), the top middle shows the second coadded image (from images 5-8), and 
the top right shows the reference image, all zoomed in to the vicinity 
of the candidate. The bottom left plot shows the first subtracted image (the 
first coadded image minus the reference), the bottom middle shows the second 
subtracted image. The bottom right panel displays information about the 
candidate, including the 
right ascension, declination, magnitude, signal-to-noise, FWHM (these 
last three quantities are calculated by comparing the reference image with the 
coadded image of the entire night), motion (this is the variation in distance 
between the first and second coadded images in units of pixels), percentage 
flux change (between the coadded image of the night and the reference image) 
and whether a candidate has been found at these coordinates before. As well 
there are links to the SIMBAD catalogue, Minor Planet Checker, 
SDSS\footnote{http://www.sdss.org/}, 
2MASS\footnote{http://www.ipac.caltech.edu/2mass/} and 
DSS\footnote{http://archive.stsci.edu/cgi-bin/dssform} to 
help decide the importance of the candidate. 
From this information, the user manually selects 
candidates of interest and lightcurves for these candidates are generated. It 
is possible to produce two lightcurves; one which includes both the transient 
and background and one which subtracts the background (estimated using an 
annulus of inner radius $\sim$6 pixels and outer radius of $\sim$14 pixels) 
away producing the lightcurve for just the transient. 

\section{Automating the Pipeline}
\label{sec:mods}

The ROTSE image processing pipeline has been used to make some significant 
discoveries of optical transients 
\citep{Kehoe:1999zc,Baykal:2005wz,Rykoff:2005es,Gezari:2008kr,Chatzopoulos:2011jc}. 
However the follow up of GW events requires processing larger numbers of images 
that is not feasible with a widget-based, user driven setup designed to handle 
one set of images at a time. For example, 
a series of commands in the IDL 
environment\footnote{http://www.exelisvis.com/language/en$-$us/productsservices/\\idl.aspx} 
are used to produce the various lists of candidates and 
their corresponding webpages. Human scanning is then required to distinguish 
candidates of astrophysical interest from those due to poor image subtraction, 
those due to minor planets, etc. Further widget-based 
commands are then needed to produce the lightcurve of each interesting 
candidate. This procedure is user intensive and time consuming. However, 
many of these steps are algorithmic, such as checking for candidates at the 
same right ascension and declination across nights, and suitable 
for automation. We have therefore written a wrapper to the pipeline 
that automates the processing of large sets of images. 
A single command now runs the complete end-to end pipeline: 
looping over image sets, finding transients, identifying transients detected 
across multiple nights, and generating lightcurves for all transients. 

Other barriers to processing large numbers of images are the need to have an 
IDL license for each instance of a running pipeline, and a pipeline 
architecture that is designed to process only a single set of events at one 
time. We have also altered the pipeline architecture to automatically create 
separate directory structures for each set of images, allow multiple instances 
of the pipeline to run simultaneously without conflict. Furthermore, we have 
removed the need for separate IDL licenses for each instance of the pipeline 
by compiling the pipeline in an IDL virtual 
machine\footnote{http://www.exelisvis.com/language/en$-$us/productsservices/\\idl/idlmodules/idlvirtualmachine.aspx}. 
Only one license is required, and only at the compilation stage. 
Combined, the change in architecture and freedom from license restrictions 
enables the processing of multiple sets of images 
simultaneously on computer clusters. We have 
written scripts for large scale processing using the \textsc{Condor}/\textsc{DAGMan} 
job management system\footnote{http://research.cs.wisc.edu/condor} for this 
purpose. The automated processing is able to perform a complete analysis, 
identifying candidates and generating lightcurves, within a few hours 
\citep{Nuttall:2012aa}. We have verified that the automated version of the 
pipeline produces lists of candidates which are identical to those produced by 
the original manual analysis. 

\subsection{Candidate Validation and Classification}
\label{subsec:postproc}

Once the automated code has produced the lightcurve information for all the 
potential candidates identified by the pipeline, a series of pass/fail tests 
are applied to each candidate. Specifically, we test whether the candidate 
appears on more than one night, whether its coordinates overlap with a known 
variable source (by querying the SIMBAD catalogue) or with an asteroid (by 
querying the Minor Planet Checker), and if the lightcurve of the 
potential candidate varies sufficiently. This last test has two 
components: a check that the lightcurve decays sufficiently 48 hours after the 
event took place, and a chi-square test to check that the candidate's 
lightcurve is not too flat. The flatness condition is 
\begin{equation}
\chi^2_\mathrm{flat} \equiv \sum_i \left(\frac{m_{i}-\bar{m}}{\sigma_{i}}\right)^{2} > 200 \, .
\label{eqn:flatness}
\end{equation}
Here $m_{i}$ is the background-subtracted magnitude of a transient in 
image $i$ with a magnitude uncertainty $\sigma_{i}$, and $\bar{m}$ is the 
average of the $m_i$ values. Candidates with $\chi^2_\mathrm{flat} \le 200$ 
are rejected.

A least-squares linear fit in $m$ 
vs.~$\log_{10}[(t - t_\mathrm{GW})/(1\,\mathrm{day})]$ is used to calculate 
the gradient of the candidate lightcurve from 48 hours to one month after 
$t_\mathrm{GW}$, the time of the event. 
We require a $\mathrm{gradient} \ge 1$ to pass this test. 
Both this threshold and the flatness threshold were selected 
through tests comparing lightcurve data from simulated transients and 
background artefacts.

The multiple-night and flatness tests are very effective at rejecting 
non-astrophysical background, particularly image-subtraction artefacts. 
We refer to these pass/fail tests collectively as the `hard' cuts in the 
analysis. Any candidate which fails one or more of the hard cuts 
is discarded. 
The decay test is seen to reduce significantly the background of 
astrophysical transients unrelated to the GW trigger while not rejecting 
simulated astrophysical transients correlated with the GW trigger 
(see Section~\ref{sec:injperf}). The specific requirement of decay 
after 48 hours is motivated models of EM counterparts for systems with 
strong GW emission, specifically kilonovae and SGRB/LGRB afterglows. 
While there are astrophysical optical transients that do not decay 
on this timescale, such as supernovae \citep{suplight:05}, the 
expected GW emission by these sources make them less likely to produce 
sources of GW triggers than compact-object mergers.

The final candidate list following application of these tests typically 
contains fewer than 5 candidates. In order to better assess the statistical 
significance of any surviving candidates, we assign to each an 
\textit{ad hoc} ranking statistic $R$ defined as 
\begin{equation}
\label{rank}
R \equiv \sum_i \Theta(18 - m_i) (18 - m_i) \times w_i \, .
\end{equation} 
Here $\Theta(x)$ is the step function and $w_i$ is a weight factor defined by 
\begin{equation}
\label{weight}
w_i
  = \left\{ \begin{array}{c l} 
      1 & \hspace{4mm} t_i - t_\mathrm{GW} < 1\,\mathrm{day} \\ 
      \left( 
        1 + \log_{10}\frac{t_i - t_\mathrm{GW}}{1\,\mathrm{day}}
      \right)^{-a} & \hspace{4mm} t_i - t_\mathrm{GW} \ge 1\,\mathrm{day} 
    \end{array} \right.
\end{equation}
Here $t_\mathrm{GW}$ is the time of the GW trigger and $t_i$ is the 
time of image $i$. The power law index $a$ is chosen to be 3 as 
shown in Figure \ref{fig:model}
and magnitude 
18 is the approximate limit at the center of the FOV for the majority 
of the ROTSE images we are analysing. Candidates 
with magnitude $m_i>18$ are likely to be processing artefacts, so the $\Theta$ 
factor ensures a rank of zero for those cases. While equation (\ref{rank}) 
is \textit{ad hoc}, it has the desirable property of favouring brighter 
candidates which appear in multiple images close in time to the GW trigger.

Candidates that survive the hard cuts are looked at further in two ways. 
Firstly we see whether the candidate's coordinates overlap (to within three 
times the size of the major diameter) with a known galaxy. We use 
the Gravitational Wave Galaxy Catalogue \citep{White:2011qf}, considering only
galaxies within 50 Mpc, as this is approximately the maximum range of current 
GW detectors to NS-NS and NS-BH binaries \citep{Abbott:2011ys}. 
Secondly, we perform a chi-square 
test comparing the candidate's lightcurve with several theoretical
models: kilonovae, SGRB afterglows, and LGRB afterglows (see Section 
\ref{sec:injperf}). Candidates 
that fulfil any 
of these conditions are highlighted in the final candidate list, 
but the ranking is not altered.

\subsection{Simulated Transients \& Detection Efficiency}
\label{subsec:injscript}

Adding simulated transients (`injecting') into the ROTSE images is key to 
quantifying both the detection efficiency and the magnitude limit of the 
pipeline. 

To begin, the user selects a number of real stars from the image as model 
stars. These stars must be sufficiently bright and isolated, so that the 
injection code does not take into account the flux of any unwanted stars and 
is able to accurately determine the point spread function (PSF) of the model 
star. We note that simple models for the PSF (e.g. a Gaussian) are not 
applicable for wide FOV images such as those from 
ROTSE, as the PSF varies across the image. An injection is performed by 
selecting a random position within 100 pixels of the model star, and 
selecting the distance to the source. The flux of the 
model star (minus the background) is scaled to follow the desired 
lightcurve, such as the kilonova or afterglow models discussed in Section 
\ref{sec:injperf}. The magnitude required in each 
image is calculated by taking into account the time between the GW trigger and 
the image being taken; for our tests we assume an interval of 0.5 days elapsed 
between the trigger time and the first image \citep{White:2012am}.

\begin{figure*}
\centering
\includegraphics[width=0.45\textwidth]{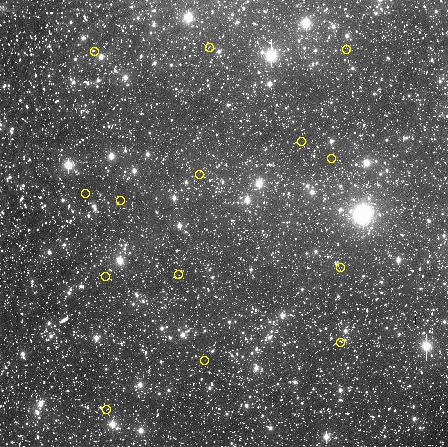} 
\includegraphics[width=0.45\textwidth]{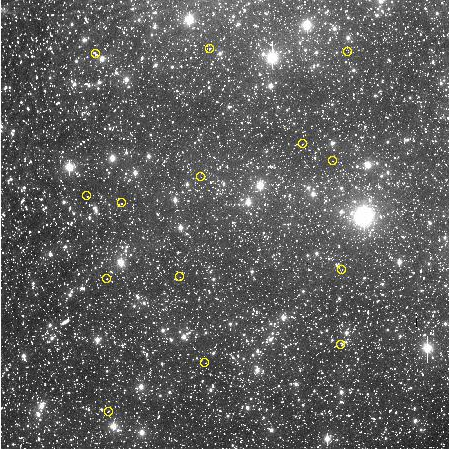}
\includegraphics[width=0.45\textwidth]{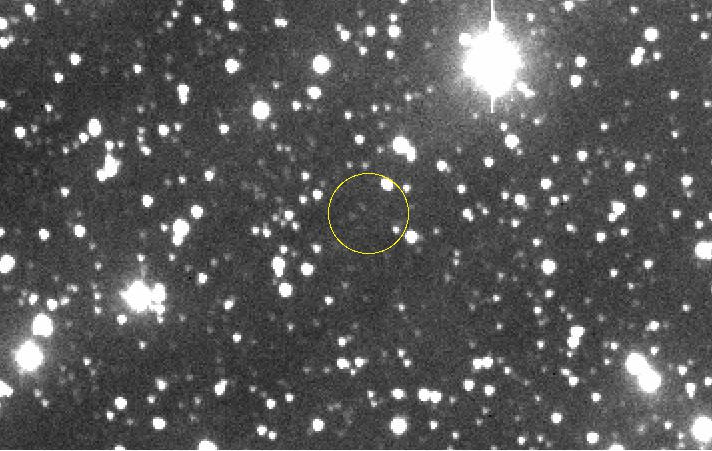} 
\includegraphics[width=0.45\textwidth]{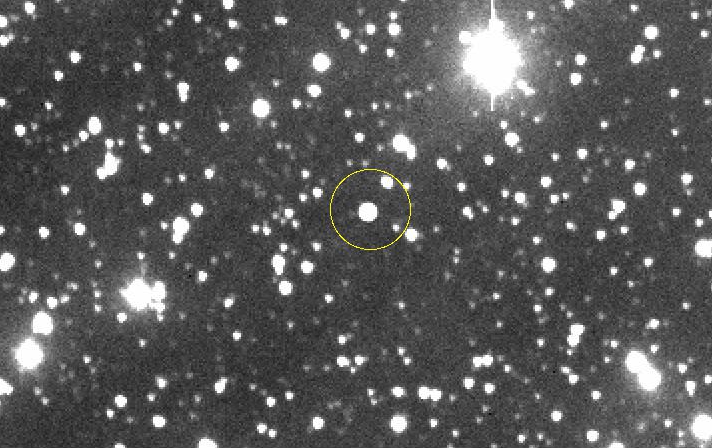} 
\caption{\label{fig:injection_mine} An example of injecting a number of 
transients into an image: 
(top left) original image; (top right) same 
image with 14 injections. The regions where the injections occurred are 
highlighted by yellow circles in both images for comparison. (bottom) 
Same images as top, focussed on the region around a single injection.}
\end{figure*}

It is vital to inject a transient not only with the correct parameters, but 
also with the correct background. 
Since the processing uses image subtraction to remove the 
background, the variation in background around the transient 
in question has to be taken into account to realistically inject a 
simulated transient into the ROTSE images.
Simply copying a model star to a new 
location in the image would produce a background around the injection 
that is significantly higher than elsewhere in the image, as the 
post-injection background would comprise both the pre-injection background at 
that location and the background around the original model star. This could 
lead to the image processing pipeline identifying fainter injected transients 
than is realistic. We therefore scale the background around the injection 
by a constant amount so that the background before and after the injection is 
comparable; see Figure \ref{fig:injection_mine} for an example.

\section{Background Study}
\label{sec:background}

Assigning a statistical significance to an event identified by the pipeline 
as associated with a GW trigger requires quantifying the false alarm 
probability. 
This is the probability of obtaining a similar event due to background, where 
for our purposes `background' includes both image-processing artefacts and 
real astrophysical transients that are not associated with a GW trigger.
To quantify this probability we have performed a background study using 
archival ROTSE data. We selected at random 102 sets of images taken in 
response to non-GW pointings over 2 years. To better mimic a GW trigger 
follow-up, each set was required to have observations spanning at least a 
month. This yielded a total of 103 sets of images. One of these was 
selected at random to be our test `GW trigger', and the other 102 were used 
for background estimation.

\begin{figure}
\centering
\includegraphics[width=0.45\textwidth]{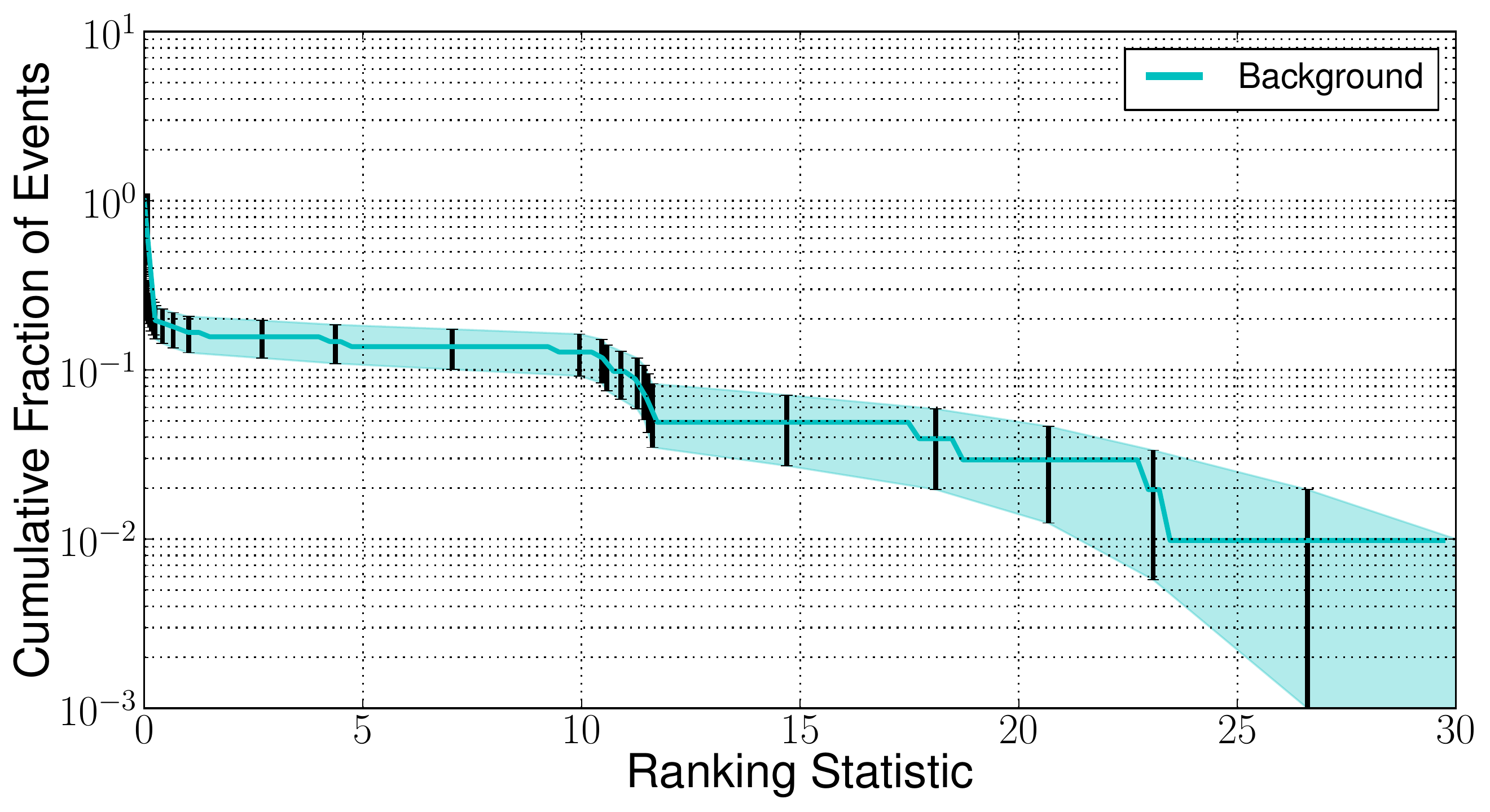}
\caption{\label{fig:background_dist}The distribution of ranking 
statistic $R$ [equation (\ref{rank})] for the highest-ranked transient in 
each of the 102 background image sets from the ROTSE archive. The error bars 
and shading indicate the Poisson errors for the background distribution.
Image sets with no candidates surviving after the hard cuts are assigned 
a rank of zero. The highest-ranked background transient over the 102 
sets has a rank of $R=29.5$.}
\end{figure}

The background is characterised as follows: each set of background 
images is processed by the automated pipeline and the highest rank 
$R$ in equation (\ref{rank}) is found. (If a background set has no 
surviving candidates after the hard cuts, a rank of zero is recorded.)
The distribution of highest-ranked events for our 102 background 
pointing sets is shown in Figure \ref{fig:background_dist}. 
We find a bi-modal distribution where approximately 
80\% of the pointings having a ranking statistic of less than 1 
and approximately 10\% have a rank greater than 11. 
The highest-ranked background event has $R\sim30$. 
A candidate in the GW trigger image set would therefore 
require $R\gtrsim11$ ($R\gtrsim30$) to have a false alarm 
probability of 0.1 (0.01) or smaller.

\section{Injection Study}
\label{sec:injperf}

Our injection studies add simulated transients corresponding to 
SGRB afterglows, LGRB afterglows, and kilonovae into the ROTSE images. 
We use the following simple models for these transients\footnote{
Equations (\ref{eqn:sgrb})--(\ref{eqn:kilonova1}) are adapted from 
\href{https://trac.ligo.caltech.edu/loocup/browser/trunk/images/catalog\_search/pipeline3/mfiles}{https://trac.ligo.caltech}
\href{https://trac.ligo.caltech.edu/loocup/browser/trunk/images/catalog\_search/pipeline3/mfiles}{.edu/loocup/browser/trunk/images/catalog\_search/pipeline3/mfiles}.
}
\begin{eqnarray}
m_\mathrm{LGRB}
  & = &  16 + \delta +\frac{8}{3}\log_{10}{\frac{t-t_\mathrm{GW}}{1\,\mathrm{day}}} 
         + 5\log_{10}\frac{D}{D_0} \, , \label{eqn:lgrb} \\
m_\mathrm{SGRB}
  & = &  23 + \delta +\frac{8}{3}\log_{10}{\frac{t-t_\mathrm{GW}}{1\,\mathrm{day}}} 
         + 5\log_{10}\frac{D}{D_0} \, , \label{eqn:sgrb} \\
m_\mathrm{kilo}
  & = &  27.9 + \frac{5}{2}\log_{10}{\frac{10^{42}\,\mathrm{erg\,s}^{-1}}{L_\mathrm{kilo}}} 
         + 5\log_{10}\frac{D}{D_0} \, , \quad \label{eqn:kilonova2} 
\end{eqnarray}
where $L_\mathrm{kilo}$ is 
the luminosity of the kilonova, 
\begin{equation}
L_\mathrm{kilo} 
  = \left\{ \begin{array}{c l} 
    10^{41.97}\,\mathrm{erg\,s}^{-1} \left(\frac{t-t_\mathrm{GW}}{1\,\mathrm{day}}\right)^{0.43} 
    & \hspace{2mm} t-t_\mathrm{GW} < 0.7 \, \mathrm{day} \\ 
    10^{42}\,\mathrm{erg\,s}^{-1} \left(\frac{1\,\mathrm{day}}{t-t_\mathrm{GW}}\right)^{1.29} 
    & \hspace{2mm} t-t_\mathrm{GW} \ge 0.7 \, \mathrm{day} \, .
  \end{array} \right. 
\label{eqn:kilonova1}
\end{equation}
Equation (\ref{eqn:lgrb}) is adapted from \citet{Kann:2007cc} and 
equation (\ref{eqn:sgrb}) from \citet{Kann:2008zg}. 
Here $\delta$ is an offset that accounts for the range in luminosities 
of GRB afterglows at fixed distance; it takes values from $\delta\simeq0$ 
for the brightest afterglows to $\delta\simeq8$ for the dimmest. 
$D$ is the distance to the source, and $D_0=6634$\,Mpc is a reference 
distance corresponding to $z=1$ (assuming 
$H_0=71\,\mathrm{km}\,\mathrm{s}^{-1}\mathrm{Mpc}^{-1}$, $\Omega_M=0.27$, 
$\Omega_\Lambda=0.73$).
$t_\mathrm{GW}$ is the time of the trigger in the 
observer frame (cosmological corrections to the time are negligible for 
all but the brightest LGRB afterglows in our analysis). 
Equations (\ref{eqn:kilonova2}) and (\ref{eqn:kilonova1}) for the kilonova 
model are adapted from \citet{Metzger:2010sy}.

Since this study there has been much work in modelling kilonovae and producing 
more realistic and comprehensive models. \citet{MeBe:12} present 
a range of plausible kilonova models which span the expected range of ejecta 
mass and velocity and also allow for realistic uncertainties in certain 
parameters. These authors specifically take in to account models put forth 
by \citet{Roberts:2011et} and \citet{Goriely:2011rp}, where the former combines 
hydrodynamic and full nuclear network calculations to determine the heating of 
ejecta material and the latter makes use of relativistic hydrodynamical 
simulations of mergers of binary neutron stars. \citet{Piran:2012wd} present 
a large set of numerical simulations which give short lived signals in the 
infrared to ultraviolet regime, powered by radioactive decay, while 
\citet{Barnes:2013wka} propose a kilonova model where the ejecta 
opacity is much higher than previously thought, leading to longer duration 
signals. In addition, progenitor models explaining the first kilonova signal 
detected in association with short GRB GRB130603B \citep{Tanvir:2013pia} have 
been proposed (see for example \citet{Hotokezaka:2013kza}).

\begin{figure}
\centering
\includegraphics[width=0.45\textwidth]{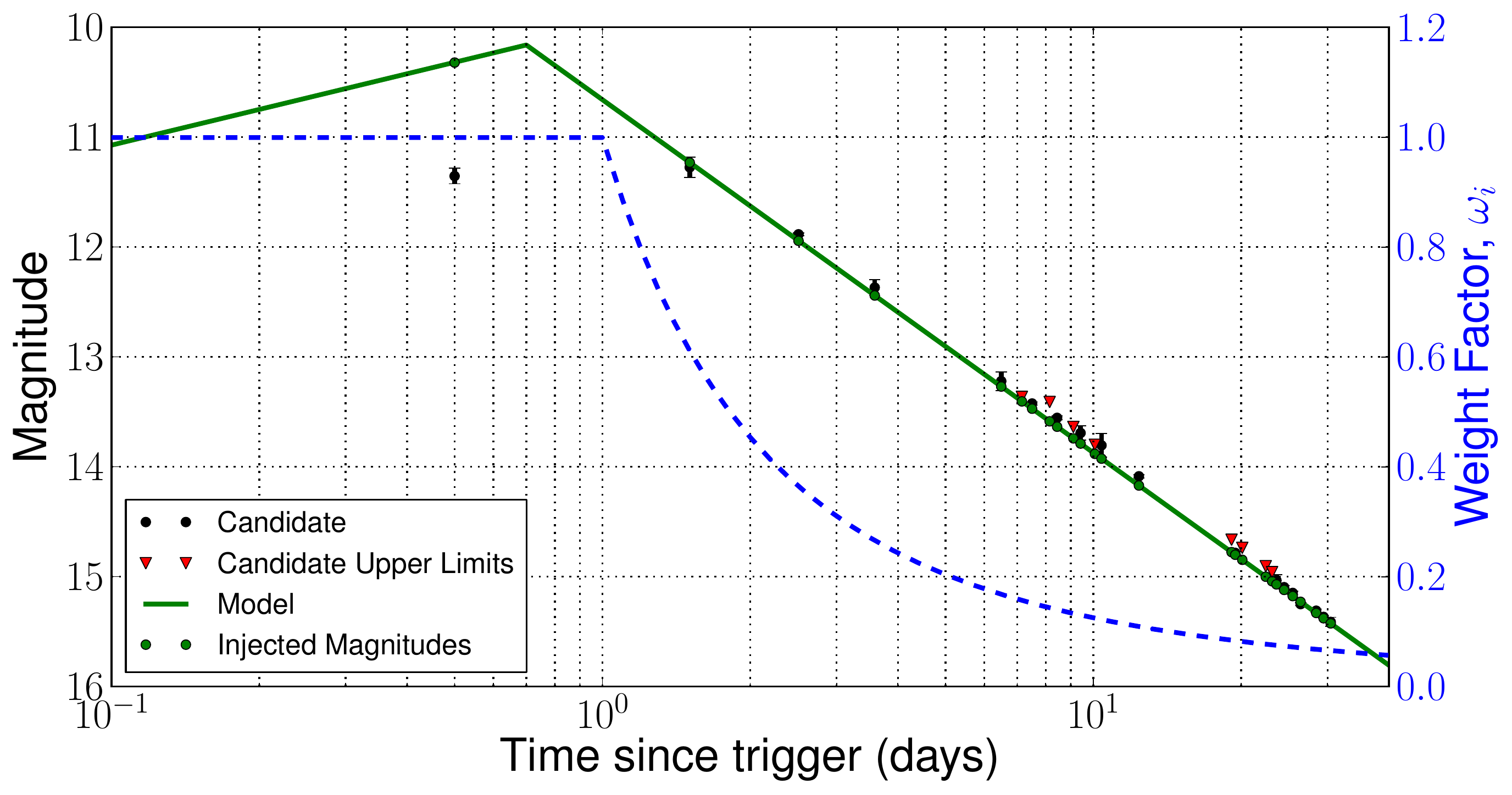}
\includegraphics[width=0.45\textwidth]{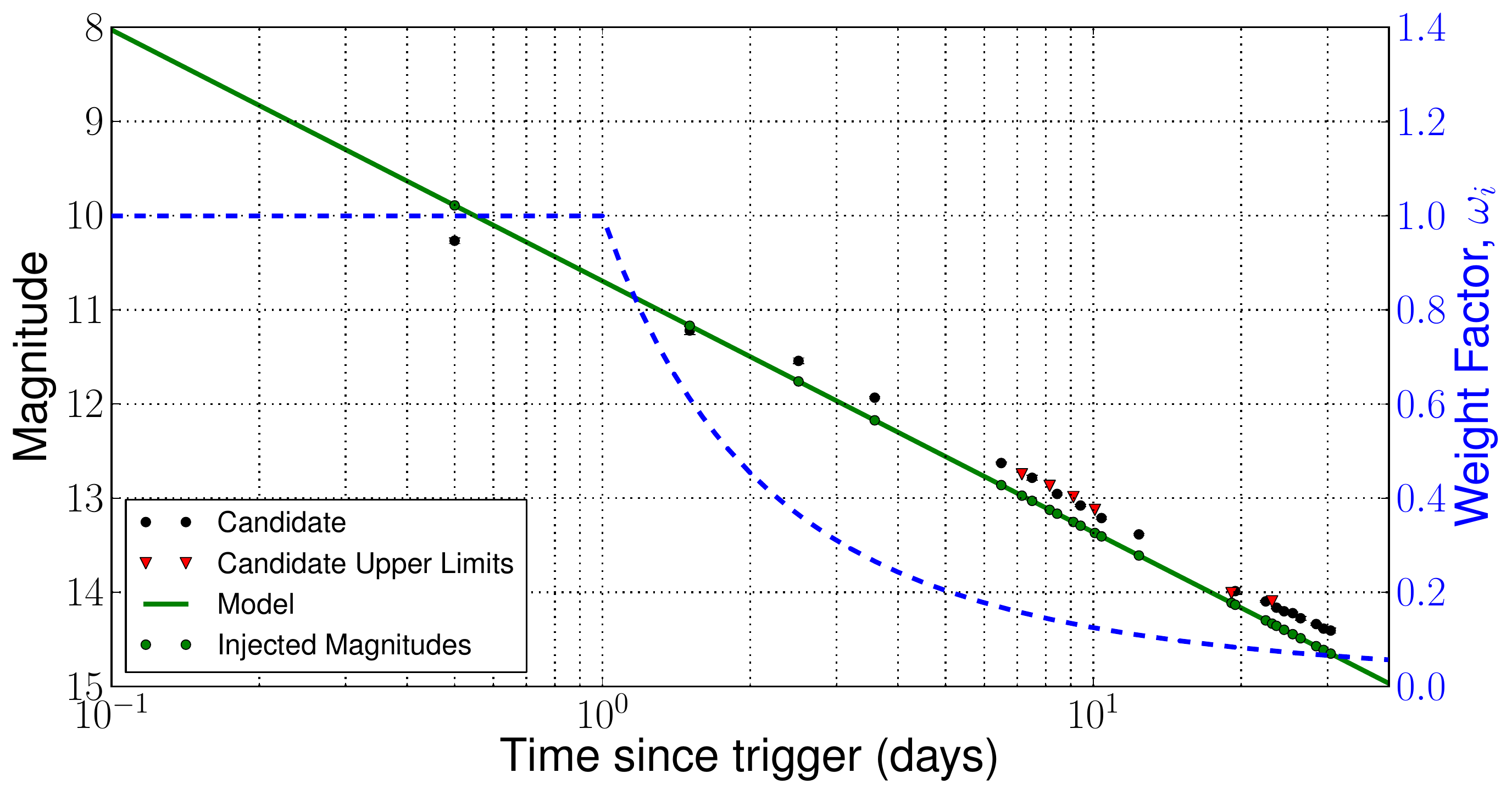}
\caption{\label{fig:model}Magnitude versus time of an injected transient
following the kilonova (top) and GRB afterglow (bottom) models. Shown are 
the transient magnitudes as reported by the automated ROTSE pipeline (black 
points) and upper limits for times when the transient was not found (red 
triangles). The magnitude of 
the injected transients are shown (green points) along with the model (green
line). For comparison the weight factor, $\omega_{i}$ [equation
(\ref{weight})], is shown by the blue dashed line.}
\end{figure}

We choose to inject the three models over a similar range of 
magnitudes, between 8 and 17 at $t=t_\mathrm{GW}+1.5$\,day. 
This corresponds to distances between 0.4 and 30 Mpc for the kilonova 
model and larger distances for the afterglow models. 
Assuming $\delta=0$ in equations (\ref{eqn:lgrb}) and (\ref{eqn:sgrb}) the 
corresponding SGRB and LGRB afterglow distances are larger by a factor of 
11 and 290, while for $\delta=8$ the distance factors are 0.28 and 7.2.
For concreteness, we assume $\delta=0$ for all distance plots. 
Example lightcurves of injected transients following the kilonova and afterglow 
models are shown in Figure \ref{fig:model}. The measured magnitudes and the 
weight factor $\omega_{i}$ [equation (\ref{weight})] are also shown for comparison.

We choose 14 reference stars in the first image as our models for the 
injections. These reference stars are spread as uniformly as possible so 
injection performance may be tested across the image. 
Each reference star is used 10 times, so that 140 injections of each 
model are performed at each distance. 

\begin{figure}
\centering
\includegraphics[width=0.5\textwidth]{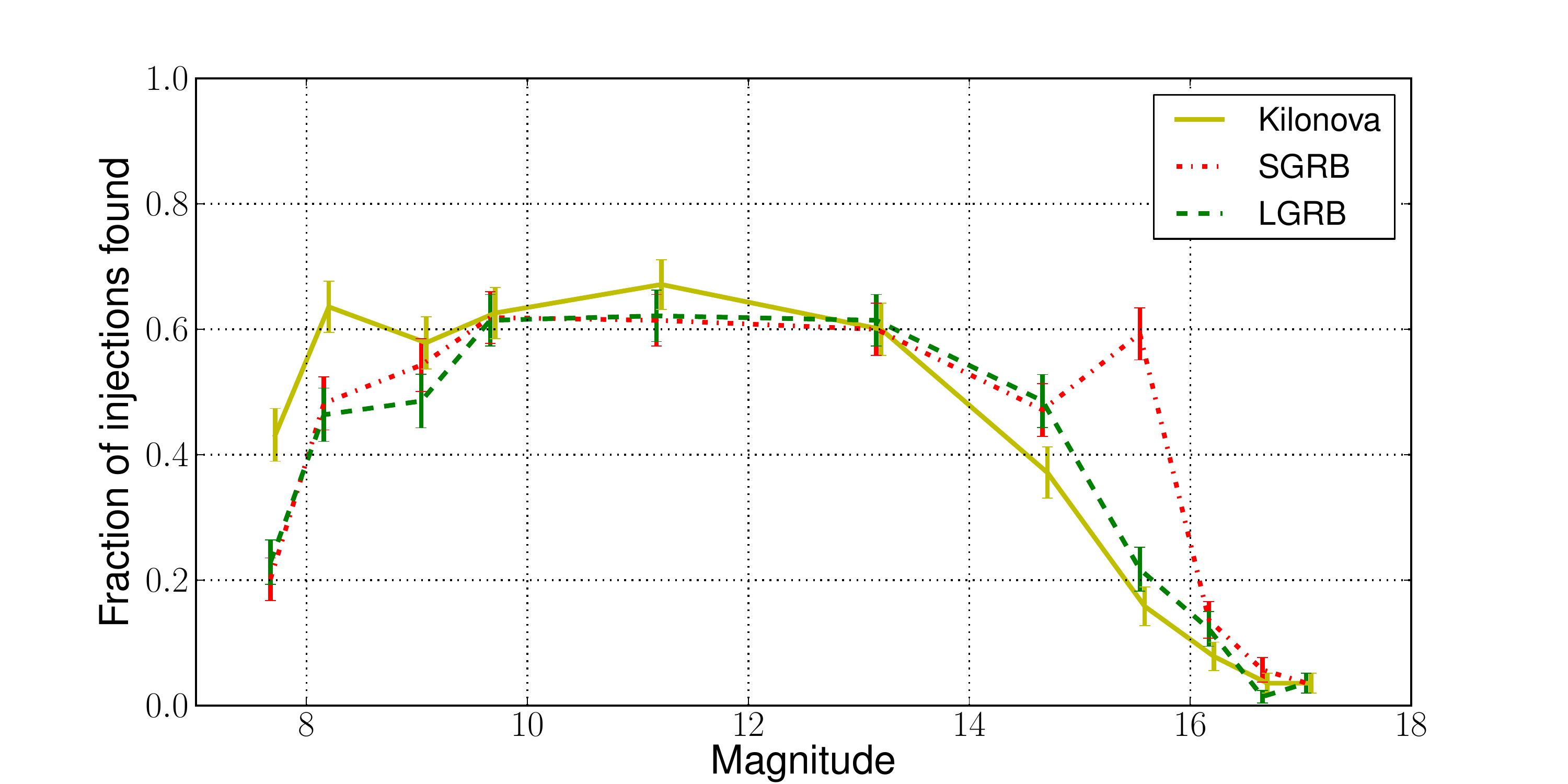}
\includegraphics[width=0.5\textwidth]{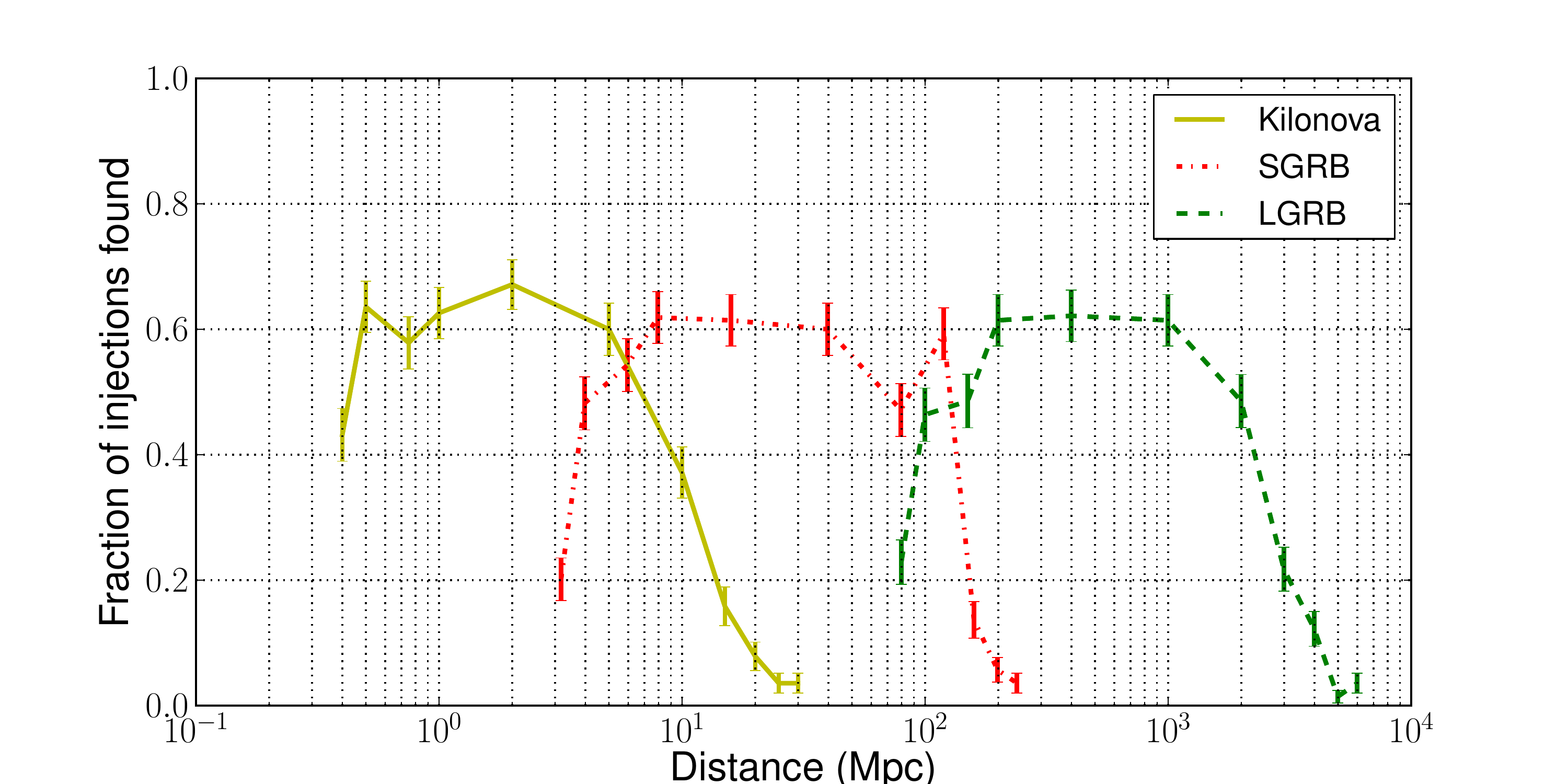}
\caption{\label{fig:efficiency}(top) Efficiency of injections found by the
automated ROTSE pipeline, with $R>0$, versus injection magnitude at 1.5 days 
after the trigger time. (bottom) Efficiency versus distance. 
The distances quoted for the GRB models assume the brightest afterglows from 
\citet{Kann:2007cc,Kann:2008zg}; i.e. $\delta = 0$ in equations (\ref{eqn:lgrb}) 
and (\ref{eqn:sgrb}). 
The worst-case luminosities ($\delta=8$) give distances a factor $10^{8/5}=40$ lower. 
All the models suffer from poor efficiency at very close distances / low magnitudes 
due to saturation, while the high-magnitude cutoff corresponds to the range of 
limiting magnitudes across the FOV. The efficiencies reach a maximum of $\sim$60\% due to poor image quality in the outer parts of the FOV, as discussed in the text.
}
\end{figure}

Figure \ref{fig:efficiency} shows the efficiency of the pipeline in 
finding injections with any rank $R>0$, in terms of distance 
and magnitude. 
The efficiency is maximum when the injection magnitude is between 
approximately 9 -- 14 at 1.5 days after the event. This is due to the 
analysis requirement that the transient be seen on multiple nights; 
since the lightcurves are decaying, the magnitude at 1.5 days tends to be 
the determining factor in whether the injection is seen on at least two nights.
The efficiency drops above magnitude 14, as this is the typical 
limiting magnitude near the edges of the image, so injections above 
this in the magnitude in the outer potions of the FOV are 
lost. The efficiency falls to zero by magnitude 17, which is the 
typical limiting magnitude of the most sensitive region at the centre 
of the FOV.

At magnitudes below 9 the injections are so bright that their image 
pixels are saturated, causing the detection efficiency to drop rapidly. 
As described in Section 
\ref{subsec:works}, the pipeline removes saturated pixels at a very early stage 
as they are assumed bad and not astrophysically 
interesting. Attempts have been made to overcome this issue by fitting each of 
the injection models to the data. The best-fit model is selected 
and used to predict the magnitude at the time of each image. 
For any images for which the candidate is not reported by the pipeline and 
for which the predicted magnitude is low enough to cause 
saturation, a new rank is calculated using the predicted magnitude for that 
time. We find that this procedure successfully retrieves transients $\sim$1 
magnitude too bright for the unaltered pipeline, but it is not effective for 
even brighter (closer) transients. However, given the distances at which 
saturation occurs this is unlikely to present a problem in practice.

The maximum detection efficiency of the automated pipeline is approximately 
60\% to 65\% for each of the models tested. 
Of the 35\% -- 40\% of injections which are not found, most are 
lost because the background-subtracted lightcurve could not be generated. 
The ability of the pipeline to produce the background-subtracted lightcurve for a transient depends on both the position in the image and on the image quality, 
as sixteen reference stars need to be identified within a 300$\times$300 pixel 
region around the transient for accurate image subtraction. 
Our ranking statistic $R$ [equation (\ref{rank})] is based on this lightcurve; 
if it is not generated then a rank $R=0$ is assigned. 
If instead we were to use the non-background-subtracted light curve, 
the peak efficiency for each model would be closer to 90\%. 

\begin{figure}
\centering
\includegraphics[width=0.5\textwidth]{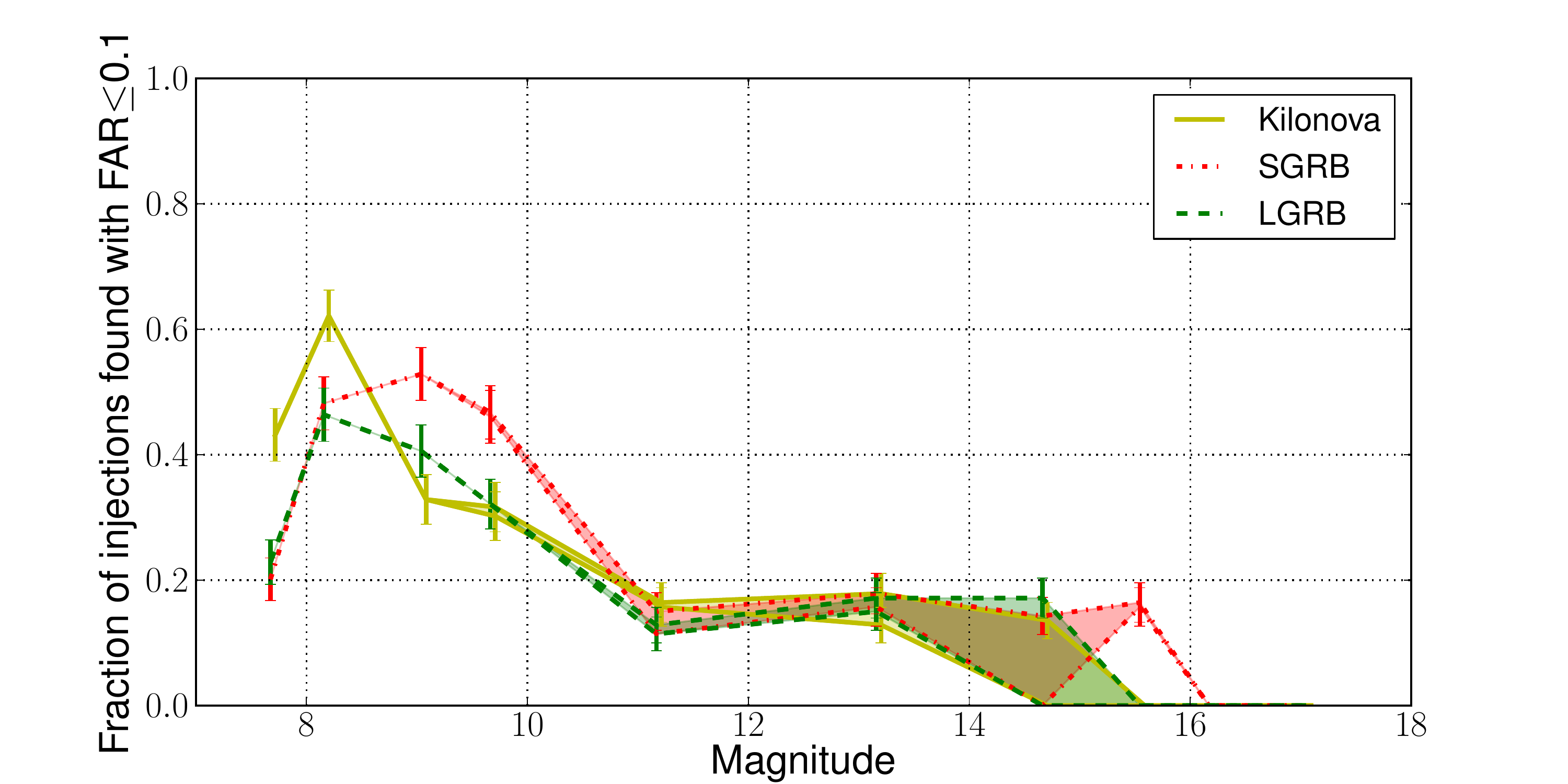}
\includegraphics[width=0.5\textwidth]{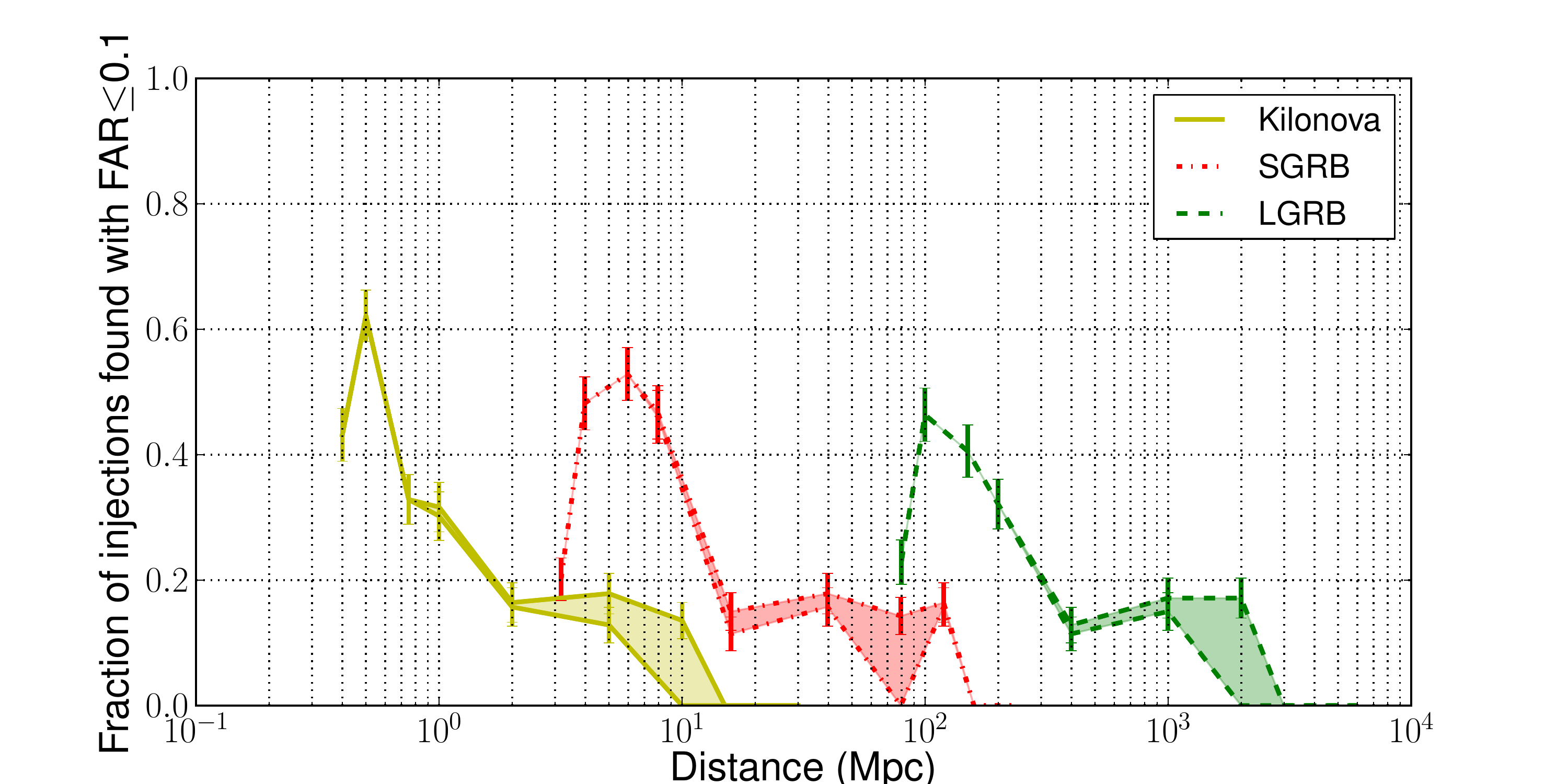}
\caption{\label{fig:efficiency_threshold}Fraction of injections found with 
rank $R>7$--$11.5$, for which the background false alarm 
probability is $<$10\%, in terms of magnitude at 1.5 days after the trigger (top)
and distance (bottom). 
The distances quoted for the GRB models assume the brightest afterglows from 
\citet{Kann:2007cc,Kann:2008zg}; i.e. $\delta = 0$ in equations (\ref{eqn:lgrb}) 
and (\ref{eqn:sgrb}). 
The worst-case luminosities ($\delta=8$) give distances a factor $10^{8/5}=40$ lower. 
The shading indicates the range of efficiencies for $R>7$ and $R>11.5$.
}
\end{figure}

\begin{figure*}
\centering
\setlength{\unitlength}{1in}
\begin{picture}(6.1,8.0) 
\put(0,6.4){\includegraphics[width=0.40\textwidth]{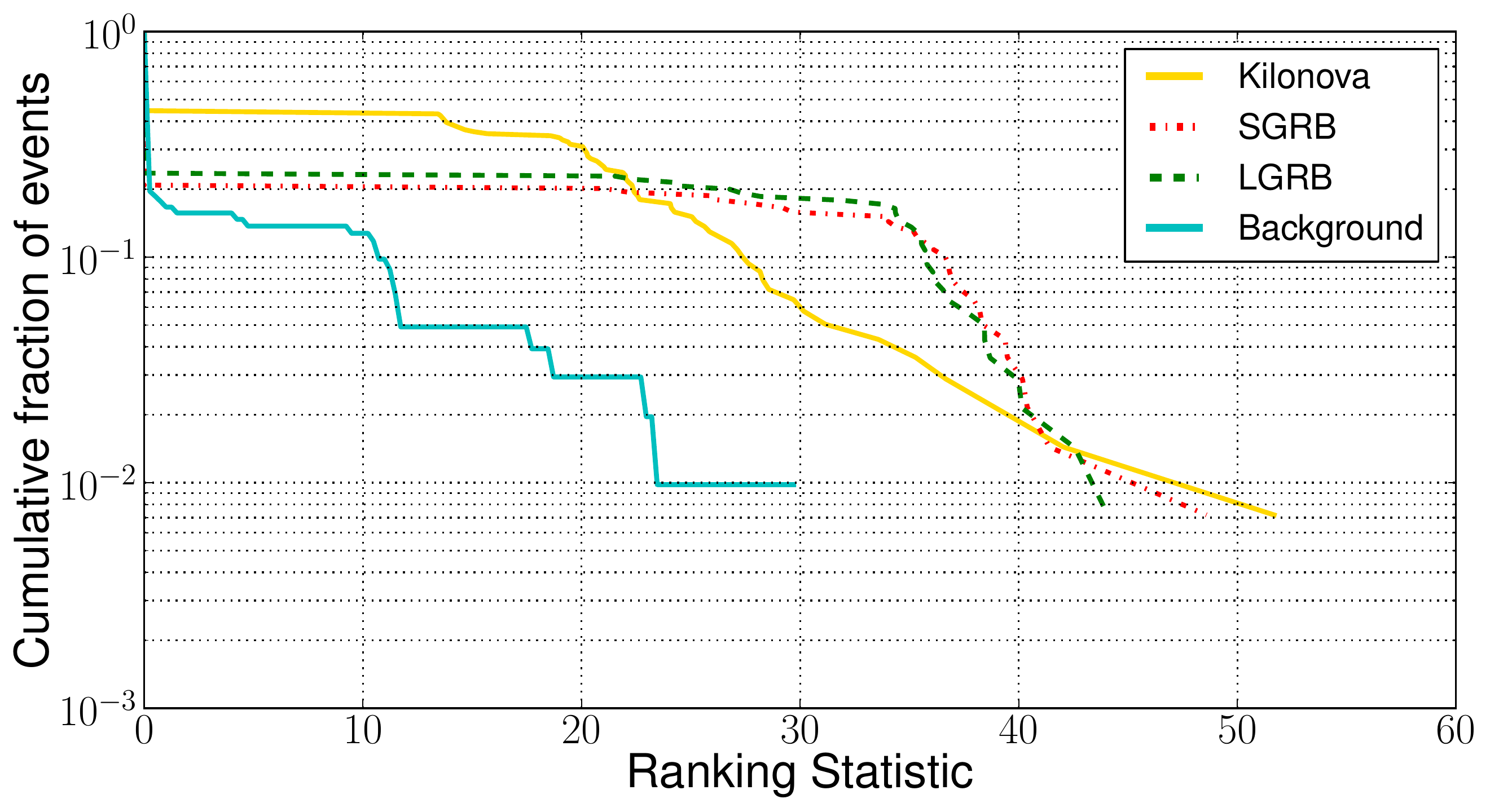}}
\put(3.1,6.4){\includegraphics[width=0.40\textwidth]{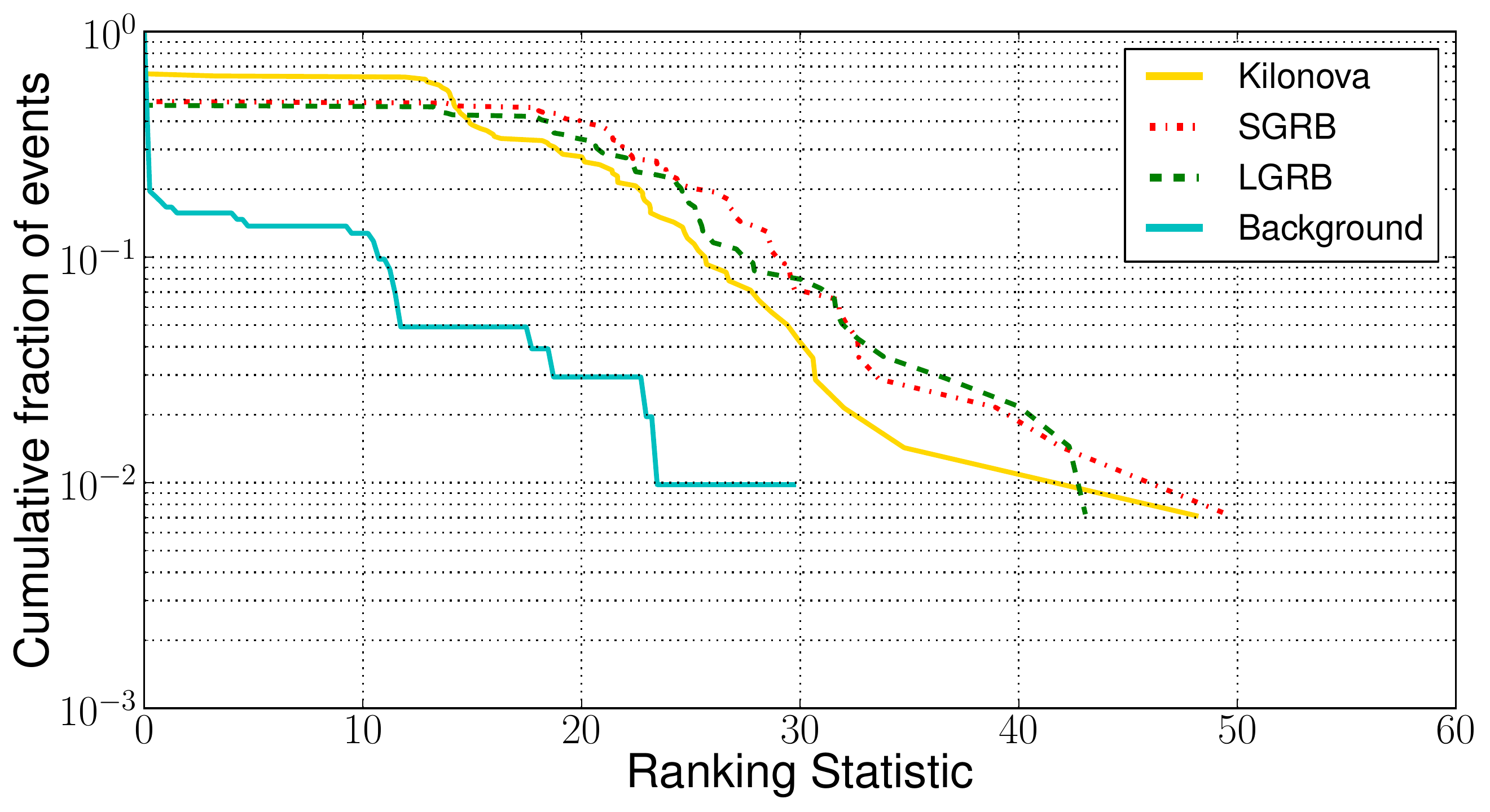}}
\put(0,4.8){\includegraphics[width=0.40\textwidth]{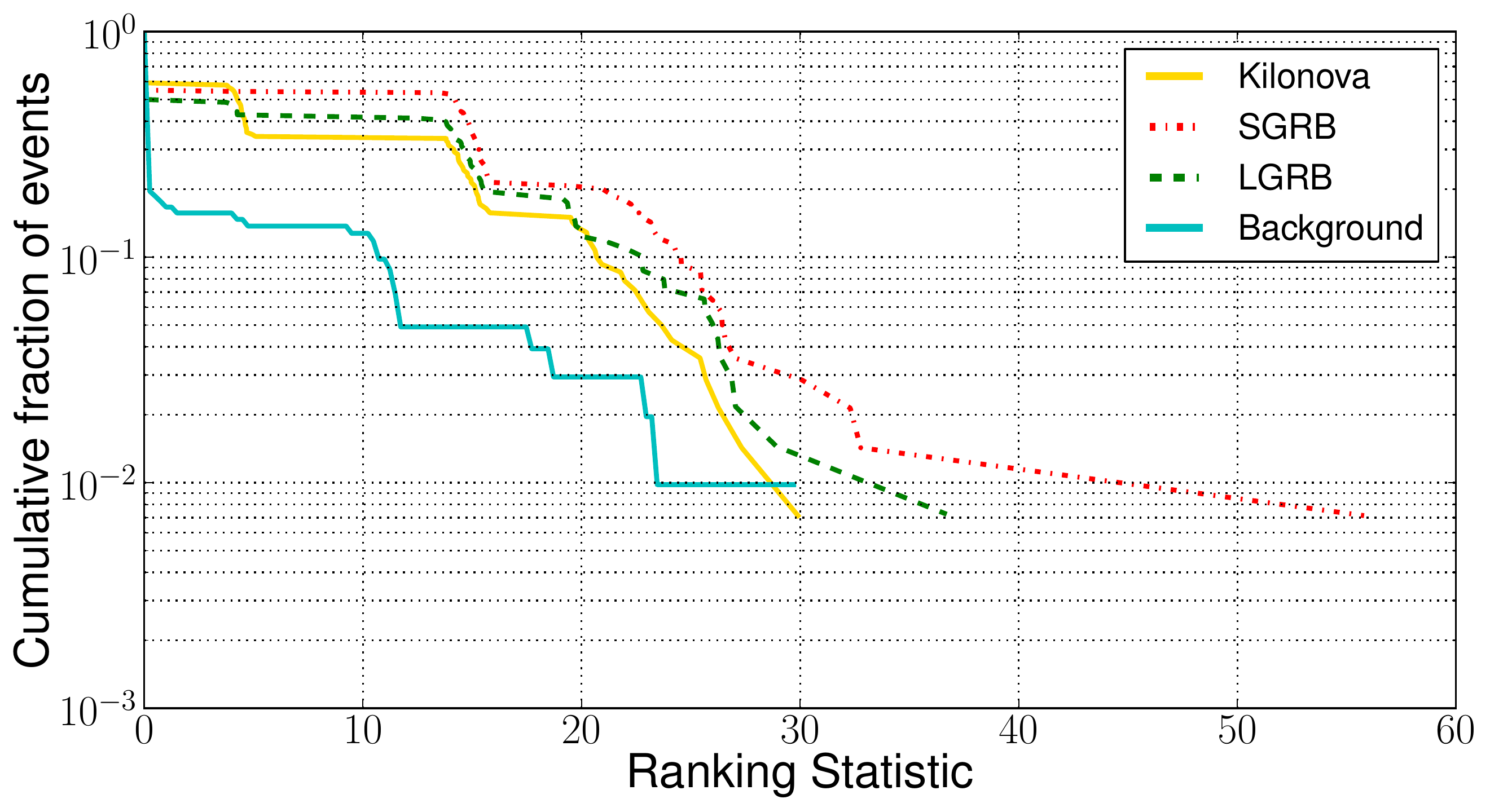}}
\put(3.1,4.8){\includegraphics[width=0.40\textwidth]{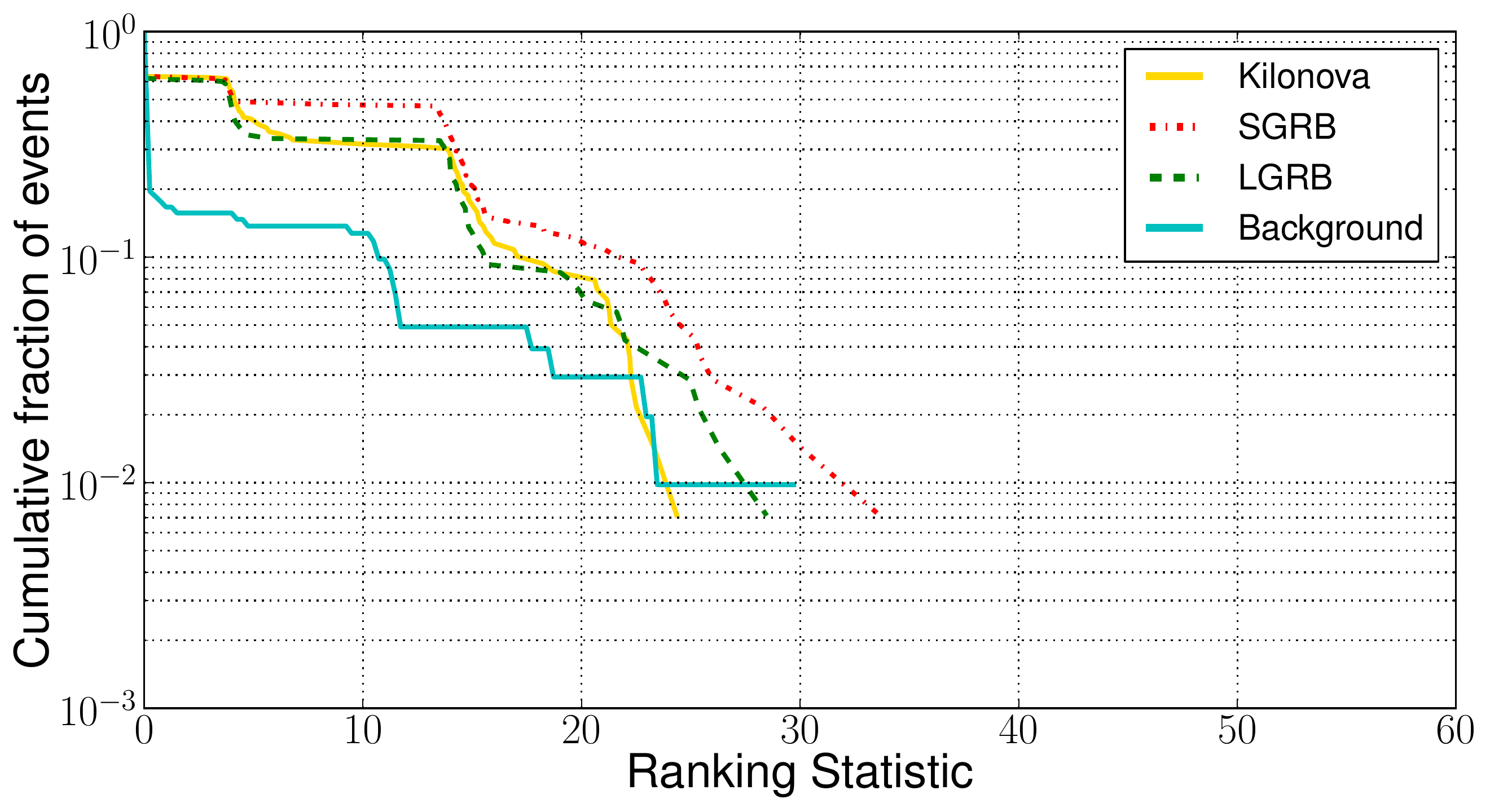}}
\put(0,3.2){\includegraphics[width=0.40\textwidth]{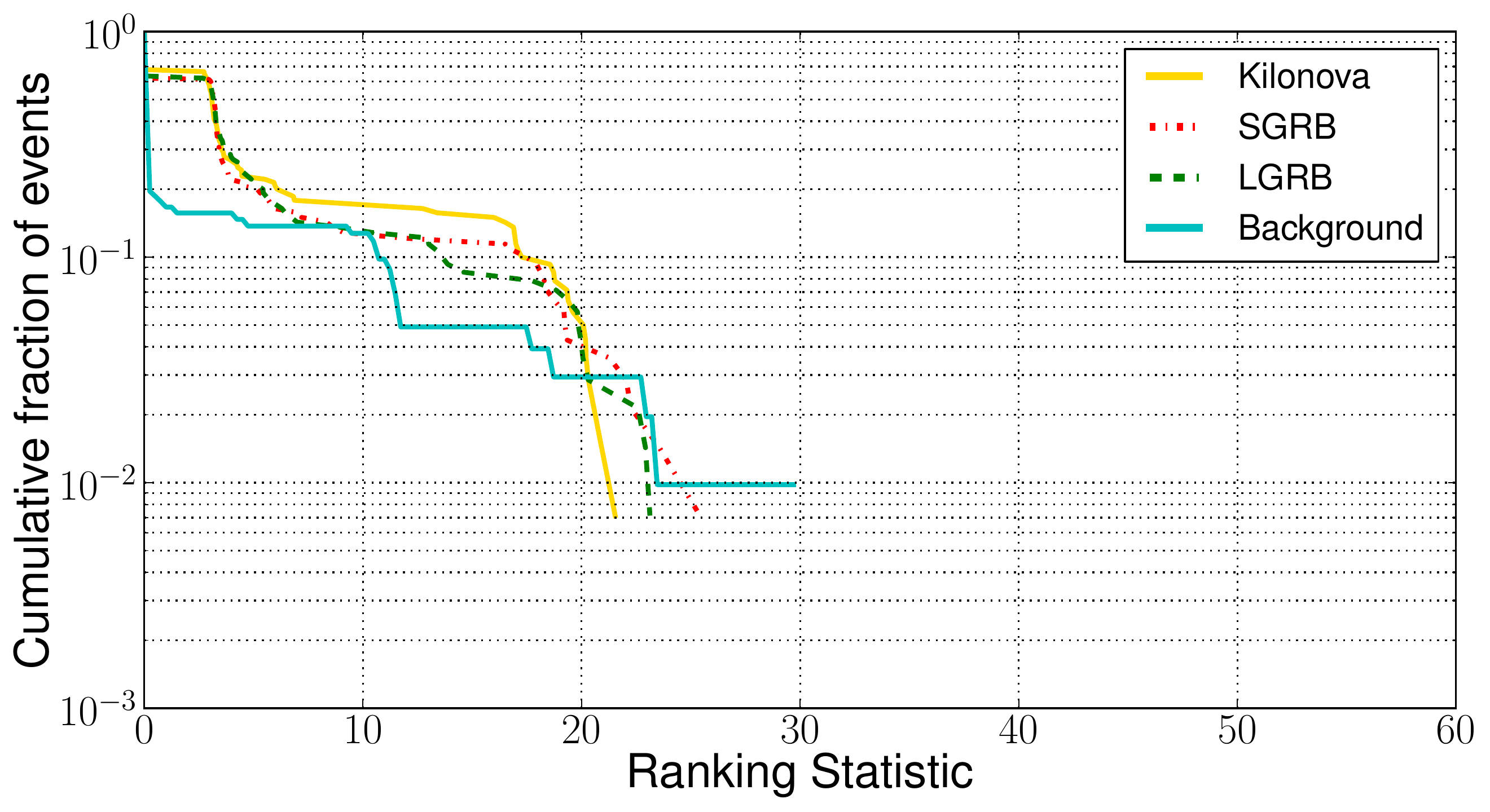}}
\put(3.1,3.2){\includegraphics[width=0.40\textwidth]{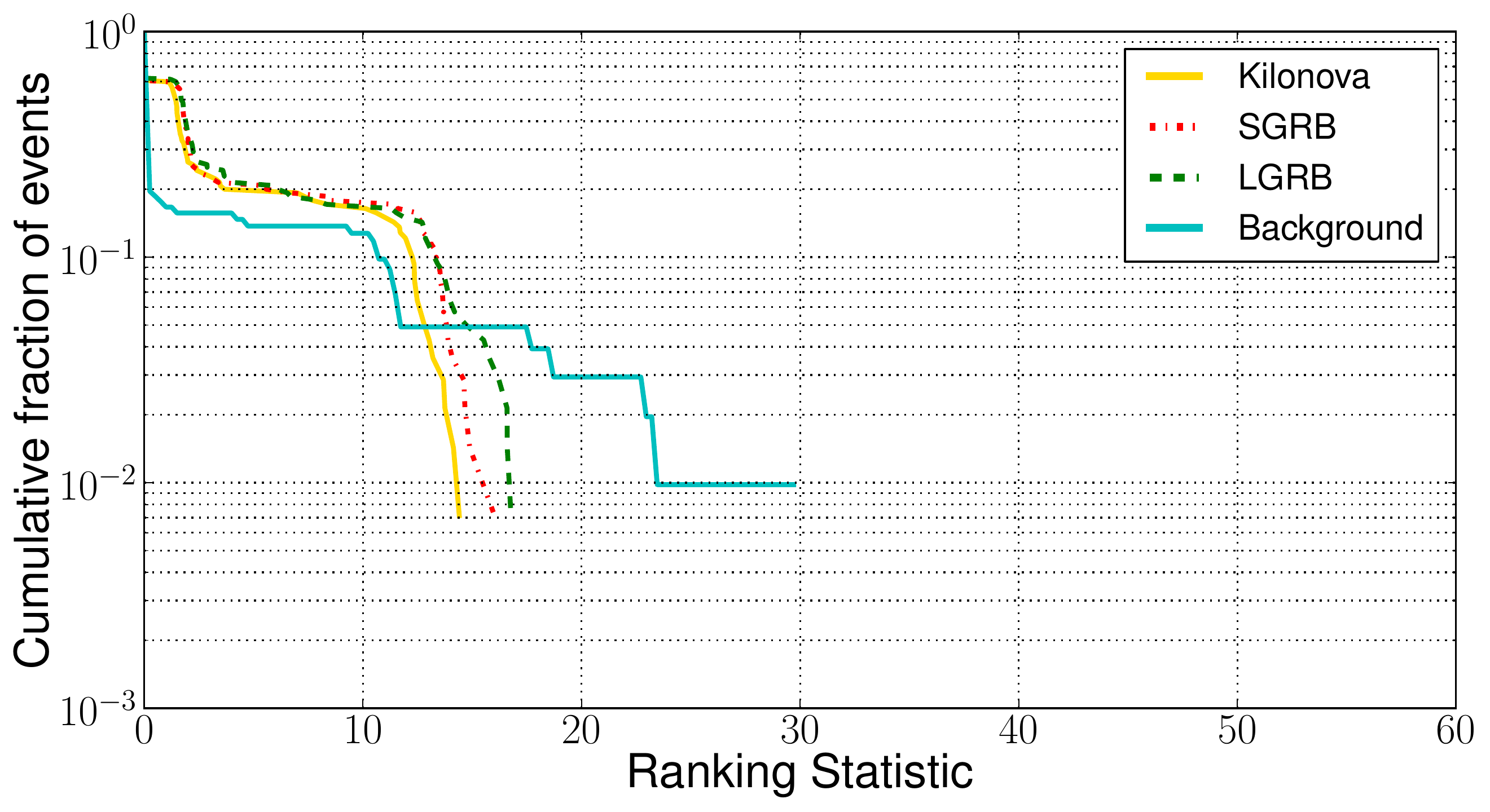}}
\put(0,1.6){\includegraphics[width=0.40\textwidth]{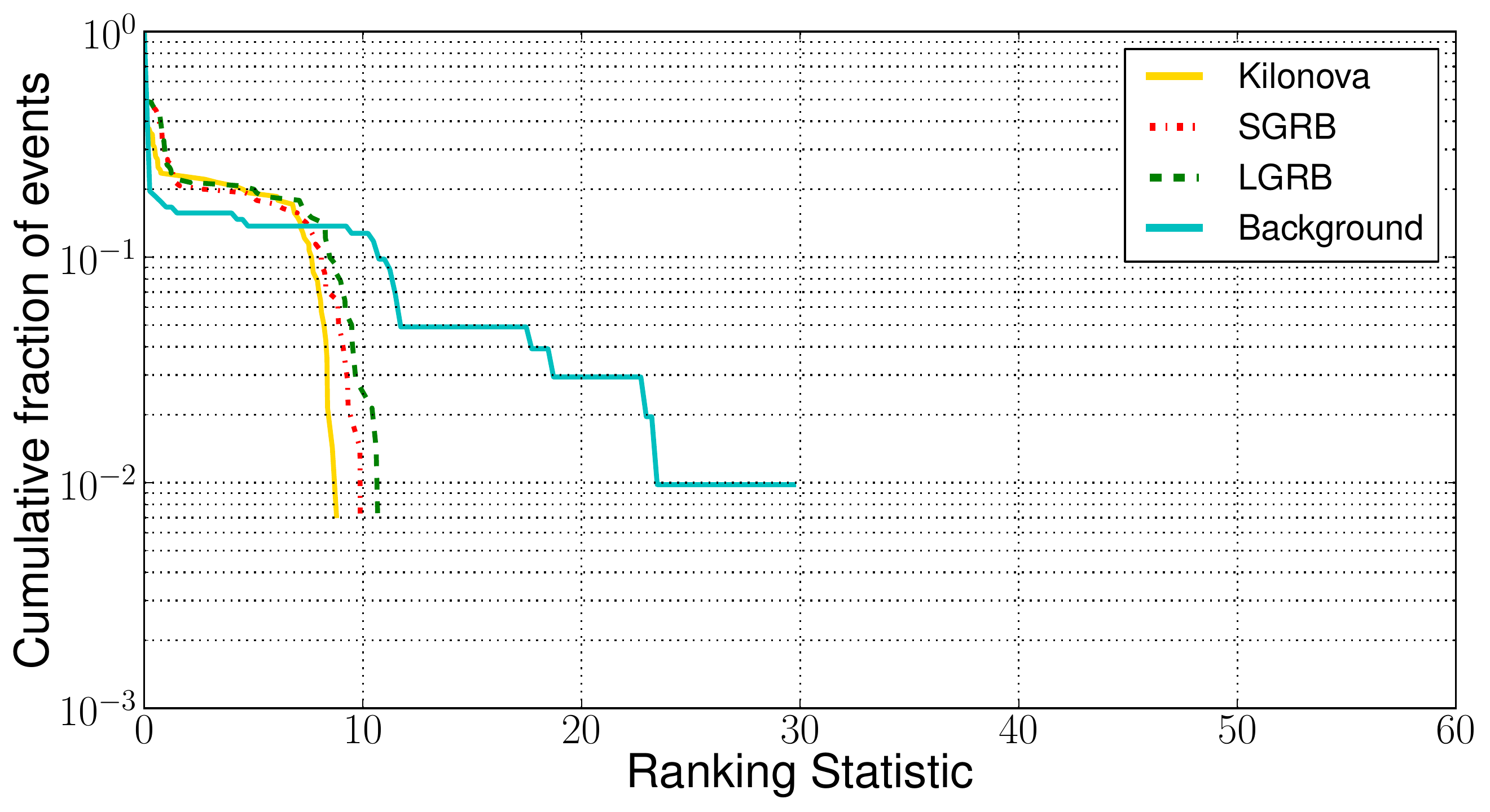}}
\put(3.1,1.6){\includegraphics[width=0.40\textwidth]{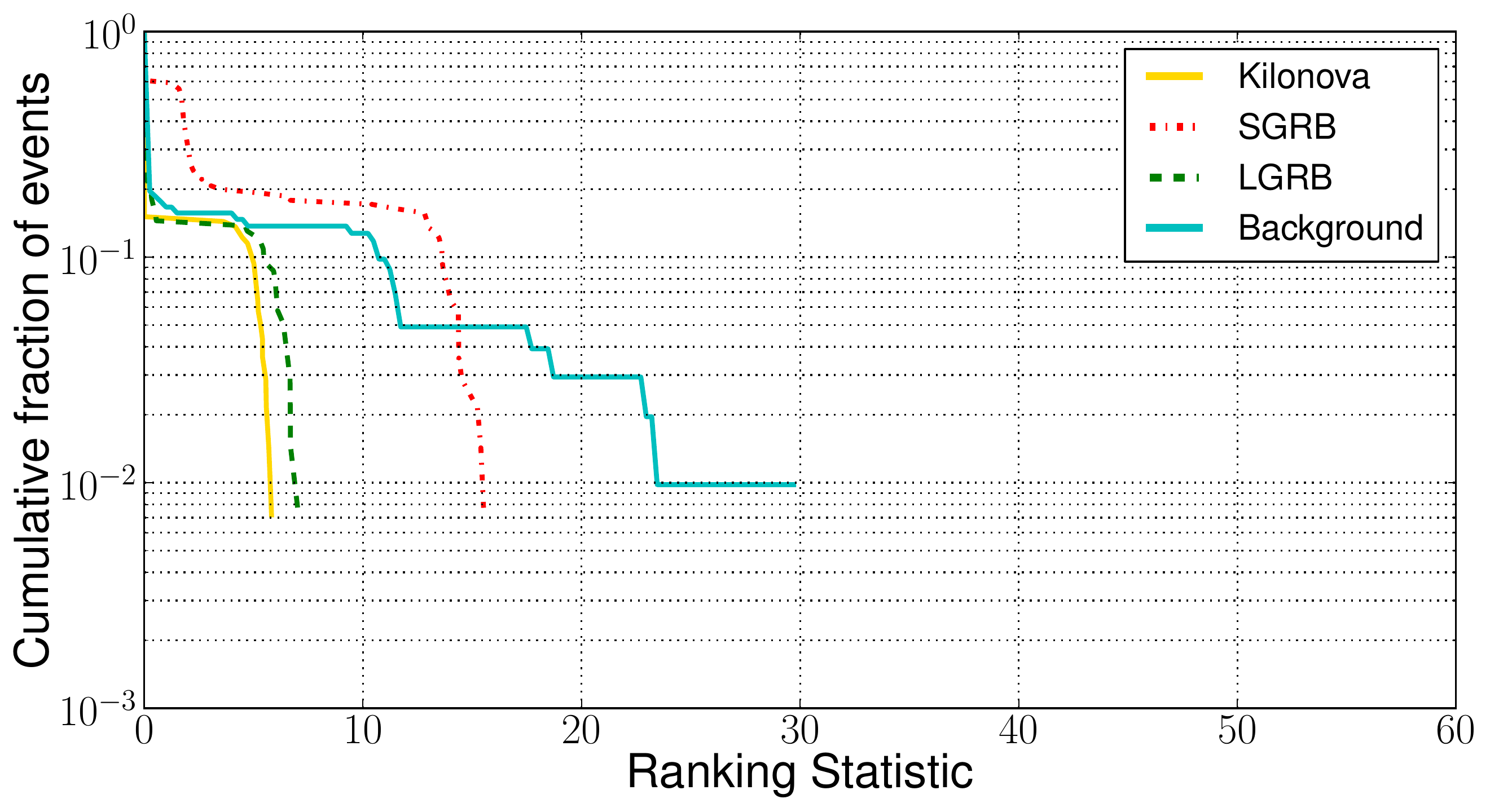}}
\put(0,0){\includegraphics[width=0.40\textwidth]{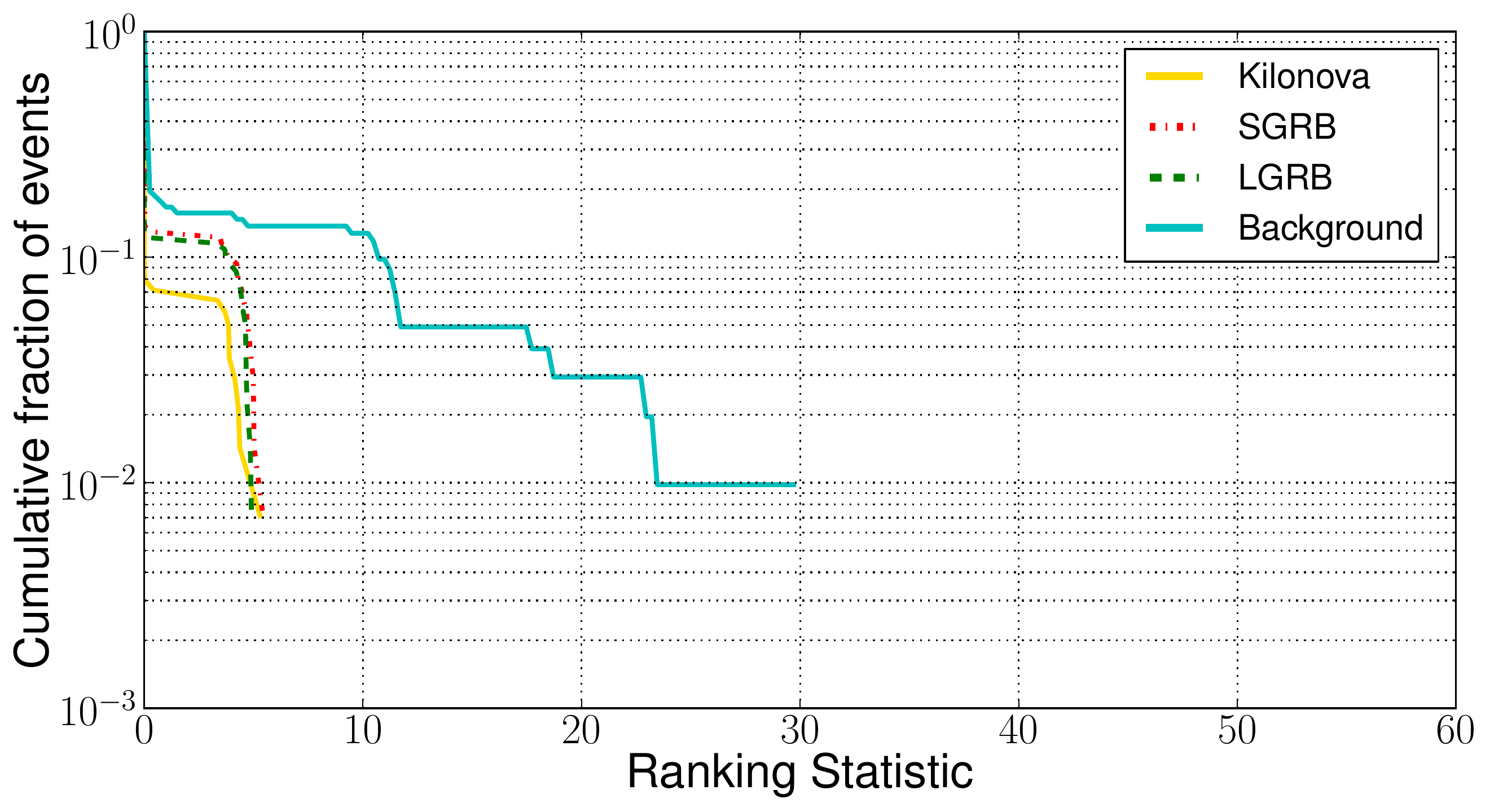}}
\put(3.1,0){\includegraphics[width=0.40\textwidth]{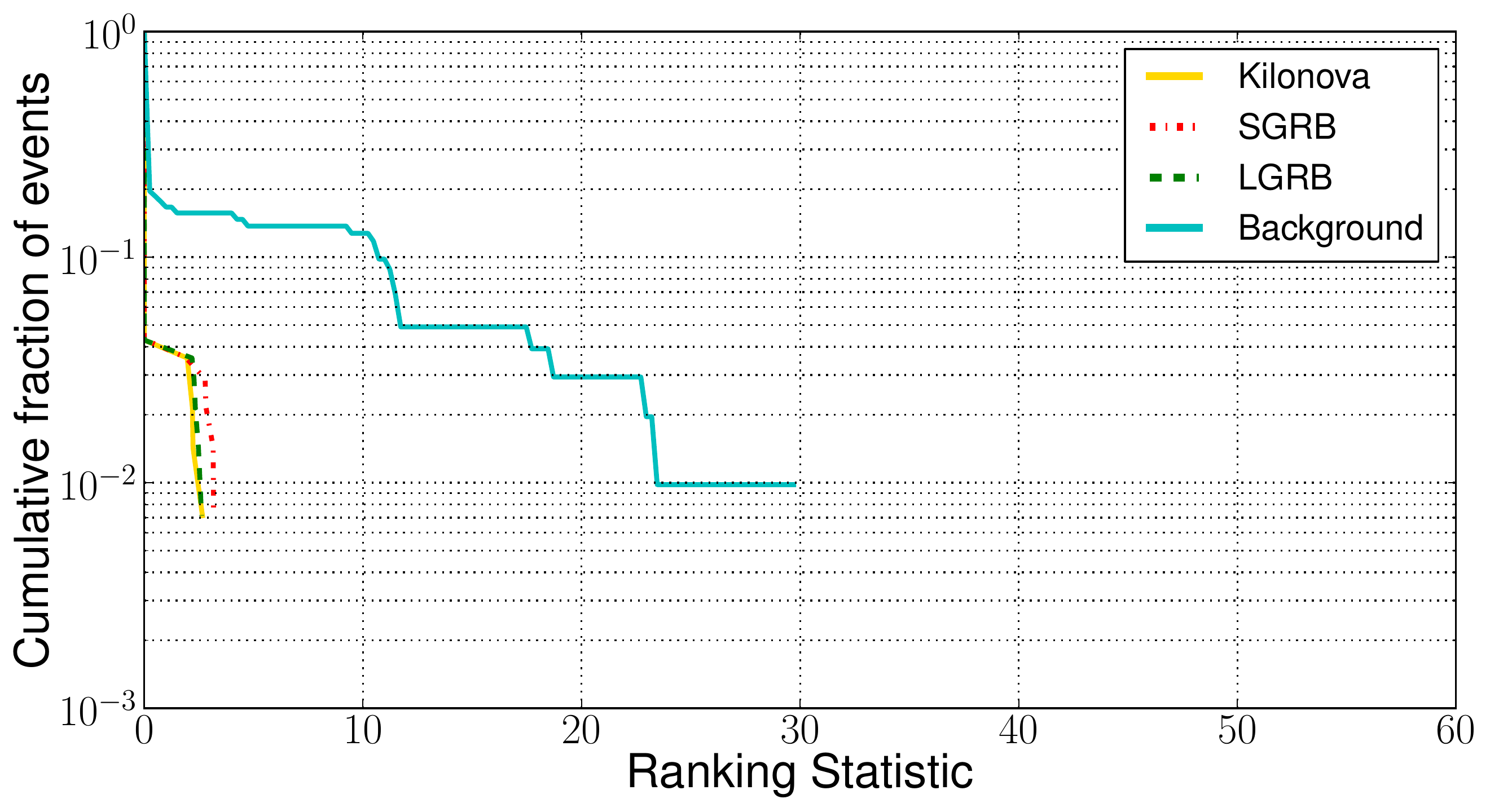}}
\put(0.35,6.65){{\bf [0.4 Mpc]}}
\put(0.35,5.05){{\bf [0.75 Mpc]}}
\put(0.35,3.45){{\bf [2 Mpc]}}
\put(0.35,1.85){{\bf [10 Mpc]}}
\put(0.35,0.25){{\bf [20 Mpc]}}
\put(3.45,6.65){{\bf [0.5 Mpc]}}
\put(3.45,5.05){{\bf [1 Mpc]}}
\put(3.45,3.45){{\bf [5 Mpc]}}
\put(3.45,1.85){{\bf [15 Mpc]}}
\put(3.45,0.25){{\bf [30 Mpc]}}
\end{picture}
\caption{\label{fig:inj_distribution}Distribution of ranking statistic values 
for injections and background for each injection distance. 
The distances quoted are those for the kilonova injections. 
The corresponding distances for the SGRB (LGRB) injections are larger 
by a factor $11\times10^{-\delta/5}$ ($290\times10^{-\delta/5}$) where $\delta$ 
is the magnitude offset in equations (\ref{eqn:lgrb}) and (\ref{eqn:sgrb}).
}
\end{figure*}

The efficiencies shown in Figure \ref{fig:efficiency} require only that the 
injection be identified with $R>0$ and pass the hard cuts; no specific 
false-alarm probability threshold has been imposed. Given the large error 
boxes expected to be associated with GW triggers, it is desirable to be 
able to identify the optical counterpart with low false alarm probability. 
As an example, we show the efficiency of detecting injections with a false 
alarm probability of less than 10\%. Due to the tail in the background 
distribution in Figure \ref{fig:background_dist}, this requires a relatively 
high rank of $R>7$ -- $11.5$.
The probability of detecting injections with this rank or higher is shown in Figure 
\ref{fig:efficiency_threshold}. The efficiencies are not as high as those found 
in Figure \ref{fig:efficiency}, with maximum vales between $\sim$45\% 
and $\sim$60\% depending on the model. 
This would suggest that all candidates which pass the hard cuts 
should be looked at further to see whether they are astrophysically 
interesting. Figure \ref{fig:inj_distribution} shows the distribution of 
injections, in terms of rank, at various distances. At very close distances the rank 
of injections is higher than the loudest candidate found in the background. At 
a kilonova distance of 1 Mpc the loudest injections are comparable to the 
loudest background event. As the distance/magnitude is increased the ranks 
slowly fall to much lower numbers, making them unexceptional when compared to 
the loudest events in the background. This again lends weight that any 
candidate to make the final candidate list be further investigated for 
significance. It also makes clear the need for additional analysis cuts 
which can eliminate the tail of the background distribution seen in 
Figure \ref{fig:background_dist}.

\section{Concluding Remarks}
\label{sec:concl}

We have presented an automated pipeline for large-scale processing of images 
from the ROTSE-III telescope system, with features appropriate for searching 
for optical counterparts to gravitational-wave events. These include the 
ability to rapidly analyse large numbers of images, which is needed both for 
covering the large GW error boxes and to be able to quantitatively estimate 
the background of optical transients and analysis artefacts unrelated to the 
GW event. The pipeline also has the ability to add simulated transients 
to the images, which are used to determine the detection efficiency of the 
pipeline and to test the background-rejection steps.
We have also proposed a simple ranking scheme for potential candidates. 
The ranking scheme favours a transient which is 
seen on multiple nights, has a low magnitude, and a decaying lightcurve. 
We have demonstrated the pipeline's performance through a 
background study of more than 100 random pointings taken from the ROTSE 
archives, as well as an injection study of more than 4500 
simulated transients added to additional archival images. 
We find that the automated pipeline detects more than 50\% of transients 
with magnitudes in the range 9 -- 14 at 1.5 days after the event. 
Injections at lower magnitudes suffer from saturation, while those at 
higher magnitudes exceed the limiting magnitude of the ROTSE telescopes 
across most of the FOV.
The detection efficiency in the magnitude 9 -- 14 range is limited by 
image quality in the outer regions of the FOV preventing construction 
of the background-subtracted light curve used to characterise transients. 

The limited efficiency in the magnitude 9 -- 14 range could be addressed 
by taking multiple overlapping images of the GW error box, so that any 
transient falls within the central $\sim$2\,deg$^2$ FOV with superior 
limiting magnitude and image quality. 
(The limiting magnitude at the edge of a ROTSE image can be lower than 
at the centre by $\sim$3 magnitudes.)
This would approximately double 
the observation time required to cover a large error box. Alternatively, 
it may be possible to decrease the number of reference stars needed to 
generate the background-subtracted lightcurve, or to use the non-subtracted 
lightcurve. The main disadvantage of the latter is that the variation in the 
candidate's magnitude is not easily identified due to the background. 

Another concern is that the distribution of background transients 
(Figure \ref{fig:background_dist}) has a large tail from the $\sim20\%$ 
of background pointings with candidates surviving the `hard' cuts. 
Visual inspection indicates that most of these are image-subtraction artefacts. 
These should be identifiable by automated tests of the shape of the transient 
in the image, which look for ring or crescent shapes. 
More generally, machine-learning techniques such as those reported 
in \citet{Abbasi:2011ja} could be employed to use all of the data 
associated with a candidate for classification.
In the meantime, the presence of the background tail motivates 
human scanning of the handful of interesting candidates identified by 
the pipeline.

The advanced LIGO and Virgo detectors should commence operations in 2015 
\citep{Aasi:2013wya}. They are expected to be able to detect binary neutron 
star mergers to a typical distance of 200\,Mpc by around the end of the 
decade, with expected detection rates of 0.2\,yr$^{-1}$ -- 200\,yr$^{-1}$. 
The ROTSE limiting 
magnitude of 14 - 17 gives a maximum sensitive distance to kilonovae 
of approximately 5\,Mpc, and a similar number for the dimmest SGRB afterglows. 
We therefore expect to require a system with limiting magnitude of 
approximately 
20-25 to detect these counterparts at typical advanced LIGO / advanced Virgo 
distances \citep{Metzger:2010sy}.
The error box from GW observations will be around 10\,deg$^2$ to 100\,deg$^2$. 
For example, simple triangulation arguments indicate that 28\% of mergers 
with have 90\% error box areas of 20 deg$^2$ or less \citep{Aasi:2013wya}. 
This requires 5-10 tilings with the ROTSE FOV. 
Given the present background in the ROTSE 
analysis, we expect $\sim$1 high-rank background event in this area, 
reinforcing the need for better background suppression and continued 
human vetting of candidates.
Recently \citet{Singer:2013xha} have demonstrated the ability to search 
over an error region of this size (71 deg$^{2}$) and detect the optical 
afterglow of a long GRB. The afterglow was identified by human scanning 
of 43 candidates produced by an automated analysis, and confirmed by 
rapid multi-wavelength followups and spectroscopic classification.

The ability to process large sets of images in a matter of hours will be 
essential in the advanced gravitational-wave detector era, where GW detections 
will be a regular occurrence. Although we know of some systems likely to 
produce both GW and EM transients, there are likely to be other sources we have 
not considered. During the next few years it is vital that we build 
tools to process EM data triggered from GW events in order to maximise the 
scientific potential of gravitational-wave observations. 

\section{Acknowledgments}

The authors thank Peter Shawhan, Jonah Kanner, Eric Chassande-Mottin and 
Marica Branchesi for useful discussions. We would also like to thank 
LIGO Laboratory and Syracuse University for use of computing clusters, with 
particular thanks to Duncan Brown, Peter Couvares, Ryan Fisher and Juan 
Barayoga. We also acknowledge NSF grants PHY-1040231, PHY-1104371, and 
PHY-0600953 which support Syracuse University Gravitation And Relativity 
computing cluster (SUGAR). ROTSE-III has been supported by NASA grant 
NNX-08AV63G and NSF grant PHY-0801007.
LKN and DJW were supported by a STFC studentship and PJS was
supported in part by STFC grant PP/501991. This
document has been assigned LIGO Laboratory document
number {LIGO}-{P1200131}-{v5}.
\bibliography{references}

\end{document}